\documentclass[a4paper,UKenglish]{article}
\usepackage{standalone}
\usepackage{amsmath,amssymb,amsthm}

\usepackage{times,amsmath,amssymb,amsthm,bm,bbm,mathtools}
\usepackage{scalefnt}
\usepackage{hyperref}
\usepackage{enumitem}

\usepackage{tikz}
\usetikzlibrary{calc,shapes}

\hypersetup{
	pdftitle={Difficult instances of the counting problem for 2-quantum-SAT are very atypical},
	pdfauthor={Niel de Beaudrap},
	colorlinks,%
	citecolor=black,%
	filecolor=black,%
	linkcolor=black,%
	urlcolor=black
}

\theoremstyle{definition}

\theoremstyle{plain}
\newtheorem{theorem}{Theorem}
\newtheorem{lemma}[theorem]{Lemma}

\renewcommand\epsilon{\varepsilon}
\newcommand\e{{\mathrm e}}

\newcommand\inter\cap
\newcommand\union\cup
\newcommand\herm{^\dagger}

\renewcommand\subset\subseteq
\renewcommand\setminus\smallsetminus

\renewcommand\ge\geqslant
\renewcommand\le\leqslant

\newcommand\idop{\mathbbm 1}

\newcommand\half{\mbox{1\!\!\:/2}}

\makeatletter
	\newcommand\ket[1]{
			\left\lvert#1%
			\@ifnextchar\bra{\right\rangle\mspace{-5mu}}{%
				\@ifnextchar\bracket{\right\rangle\mspace{-5mu}}{\right\rangle}}%
}
\newcommand\bracket[2]{%
	\left\langle#1\mspace{-3mu}\left|#2%
	\@ifnextchar\bra{\right\rangle\right.\mspace{-5mu}}{%
		\@ifnextchar\bracket{\right\rangle\right.\mspace{-5mu}}{\right\rangle\right.}}%
	}
\makeatother
\newcommand\bra[1]		{\left\langle#1\right\rvert}		

\newcommand\ens[1]		{\left\{ #1 \right\}}

\renewcommand\vec\mathbf

\makeatletter

\newcommand\ox\otimes
\newcommand\x\times

\DeclareMathOperator\poly{poly}

\newcommand\Z{\mathbb Z}

\newcommand\C{\mathbb C}

\newcommand\cH{\mathcal H}
\newcommand\cC{\mathcal C}

\DeclareMathOperator*\bE{\mathbb{E}}
\DeclareMathOperator*\Var{Var}

\newcommand\sur[1]{^{(#1)}}

\newcommand\arc{\rightarrow}

\newcommand\num{\problem\#}

\renewcommand\P{\ensuremath{\mathsf P}}

\newcommand\NP{\ensuremath{\mathsf{NP}}}

\newcommand\problem[1]{\mbox{\bfseries\upshape{\scalefont{0.85}{#1}}}}
\renewcommand\problem[1]{\textsc{\lowercase{#1}}}
\renewcommand\problem[1]{\mbox{\textup{\scalefont{0.85}{\uppercase{#1}}}}}

\newcommand\CNFSAT[1][]{\ensuremath{#1}\problem{-CNF-SAT}}
\newcommand\XORSAT[1][]{\ensuremath{#1}\problem{-XOR-SAT}}
\newcommand\SAT[1][]{\ensuremath{#1}\problem{-SAT}}
\newcommand\MAXSAT[1][]{\problem{MAX-}\ensuremath{#1}\problem{-SAT}}
\newcommand\QSAT[1][]{\ensuremath{#1}\problem{-QSAT}}

\newcommand\parStyle[1]{\textrm{\mdseries\upshape({#1}\kern0.1ex)}}
\newenvironment{romanum}{%
	\begin{enumerate}[label=\parStyle{\itshape\roman*},labelwidth=\romanumlabelwd,leftmargin=1\romanumlabelwd,itemsep=0.2ex]%
}{%
	\end{enumerate}%
}
\newlength\romanumlabelwd
\newcommand\parit[1]{\textup{(\textit{#1})}}

\makeatletter
\newcommand\eg{\emph{e.g.}}
\newcommand\ie{\emph{i.e.}}
\newcommand\iid[1]{\emph{i.i.d}\@ifnextchar.{}{.#1}}
\newcommand\etc{\@ifnextchar.{\emph{etc}}{\emph{etc.}}}
\newcommand\etal{\@ifnextchar.{\emph{et al}}{\emph{et al.}}}
\makeatother

\makeatletter
\renewcommand\paragraph{\@startsection{paragraph}{4}{\z@}%
                                    {3.25ex \@plus1ex \@minus.2ex}%
                                    {-1em}%
                                    {\normalfont\normalsize\itshape}}
\makeatother

\newcommand\ErdRen{{Erd\H{o}s}--{R\'enyi}}

\newif\ifpreliminaries		\preliminariestrue
\newif\ifscalingregimes		\scalingregimestrue
\newif\ifprelimanalysis		\prelimanalysistrue
\newif\ifpreamble					\preamblefalse

\begin{document}
\title{%
	\vspace{-9mm}%
	Difficult instances of the counting problem for 2-quantum-SAT are very atypical}
\author{Niel de Beaudrap\thanks{%
	 Supported by a Vidi grant from the Netherlands Organisation for Scientific Research (NWO) and the European Commission project QALGO.}\\
	 \textsf{\normalsize beaudrap@cwi.nl}\\
	CWI, Science Park 123, 1098 XG Amsterdam, Netherlands}
\date{1 July 2014}
\maketitle

\vspace{-4ex}
\begin{abstract}
	The problem \emph{2-quantum-satisfiability} (\QSAT[2]) is the generalisation of the \textsc{2-cnf-sat} problem to quantum bits, and is equivalent to 
	determining whether or not a spin-\half\ Hamiltonian with two-body terms is frustration-free.
	Similarly to the classical problem \num\SAT[2], the counting problem \num\QSAT[2] of determining the size (\ie~the dimension) of the set of satisfying states is \num\P-complete.
	However, if we consider random instances of \num\QSAT[2] in which constraints are sampled from the Haar measure, intractible instances have measure zero.
	An apparent reason for this is that almost all two-qubit constraints are entangled, which more readily give rise to long-range constraints.
	
	We investigate under which conditions product constraints also give rise to efficiently solvable families of \num\QSAT[2] instances.
	We consider \num\QSAT[2] involving only discrete distributions over tensor product operators, which interpolates between classical \num\SAT[2] and \num\QSAT[2] involving arbitrary product constraints.
	We find that such instances of \num\QSAT[2], defined on \ErdRen\ graphs or bond-percolated lattices, are asymptotically almost surely efficiently solvable except to the extent that they are biased to resemble monotone instances of \num\SAT[2].
\end{abstract}


\section{Introduction}

Local spin Hamiltonians are simplified models for physical systems, in which the 
system is approximated by 
finite-range interactions between particle sites in a fixed network.
We consider problems which involve the minimum eigenvalue of two-body Hamiltonians,
$	H	=	\sum_{\langle u,v \rangle} h_{u,v}	\;, $
for projectors $h_{u,v}$ acting on pairs of qubits (\ie~spin-\half\ particles) $u$ and $v$ drawn from some set $V$.
When each $h_{u,v}$ is a projector onto standard basis states, finding the minimum energy of $H$ is in effect \MAXSAT[2], or the problem of finding an assignment to boolean variables which satisfies as many constraints as possible, from a given list of constraints on pairs of bits.
Minimum eigenspace problems are therefore at least \NP-hard in general, and are even \NP-hard to approximate to within a small percentage error~\cite{PY91}.
Even if the minimum energy is known, determining the degeneracy (the dimension of the lowest-energy eigenspace) is \num\P-hard in general, or as difficult as determining the number of satisfying solutions to an instance of \SAT[3]~\cite{Valiant79}.
Thus, such problems should be considered to be intractable, barring major and unexpected advances in technique. 

This article concerns the conditions under which computing the degeneracy of local Hamiltonians on spin-\half\ particles is possible in polynomial time, as opposed to its worst-case complexity of being \num\P-hard.
We make this question more precise below. 

\subsection{Counting problems for frustration-free spin-\half\ Hamiltonians}

A special case of interest are \emph{frustration-free} Hamiltonians, for which there are states $\ket\psi$ which minimize all of the terms $\bra\psi h_{u,v} \ket\psi$ simultaneously.
Finding ground states of such systems may still be hard, but one may at least certify succinct descriptions of ground states, \eg~by direct calculation of energy contributions from each term $h_{u,v}$.
These models are therefore a potentially useful proving ground for analytical techniques in many-body theory.
Indeed, 
there is a wide class of such Hamiltonians on qubits, for which one may efficiently characterise the ground-state manifold~\cite{dBOE10}.

Bravyi~\cite{Bravyi06} defines the \emph{quantum satisfiability problem}, or \QSAT[k] (for any fixed $k \ge 1$), to be essentially the problem of determining whether a Hamiltonian consisting of a sum of projectors, each acting non-trivially on at most $k$ spin-\half\ particles, is frustration-free.
Bravyi shows that \QSAT[2] is efficiently solvable; by contrast, \QSAT[3] may not have any efficient solutions, even if it were somehow shown that $\P = \NP$~\cite{3-QSAT}.

A natural problem for frustration-free systems is to determine the ``degeneracy'' of their ground-state energy levels.
Given a two-body spin-\half\ Hamiltonian $H$ as input, let \num\QSAT[2] denote the problem of computing the dimension of the subspace of states which minimizes the energy contributions of each interaction term of $H$ independently.
We refer to this dimension as the \emph{value} of the instance of \num\QSAT[2].
This value is positive if and only if $H$ is frustration-free, and greater than one if $H$ is also degenerate.
The name \num\QSAT[2] is chosen (see also Ref.~\cite{JWZ11}) in analogy to the problem \num\SAT[2] of counting the satisfying assignments to an instance of \SAT[2] .
The dimension of the ground-state manifold of a frustration-free spin-\half\ Hamiltonian is simply the size of a basis for the solution space: if the projectors $h_{u,v}$ are all diagonal operators, this problem is \num\SAT[2].
Thus \num\QSAT[2] may be construed as a counting problem in the traditional sense.

While \SAT[2] is efficiently solvable, the counting problem \num\SAT[2] is \num\P-complete~\cite{Valiant79}, \ie~polynomial-time equivalent to counting satisfying assignments for instances of \CNFSAT[3].
As \num\QSAT[2] generalizes \num\SAT[2], the former problem is at least as hard in the worst case.
(Ji, Wei, and Zeng~\cite{JWZ11} show that in fact $\num\QSAT[2] \in \num\P$.) 
One may ask if there are broad subfamilies of \num\QSAT[2] which are considerably easier than \num\P\ to compute, and if so, whether such conditions can themselves be easily decided.

\subsection{Entanglement and worst case vs.\ typical counting complexity}

Though \num\QSAT[2] is \num\P-complete, there is a sense in which ``generic'' instances of \num\QSAT[2] are easily solved.
Fix any graph $G$ on $n$ vertices.
If we assign a qubit to each vertex, and a term $h_{u,v} = \ket{\eta_{u,v}}\bra{\eta_{u,v}}$ for each edge $uv \in G$, where $\ket{\eta_{u,v}}$ is distributed according to the Haar measure, the resulting \num\QSAT[2] instance can be easily solved (except with probability 0) from the structure of $G$~\cite{LMSS10,BMR-2010}.
Specifically, if the graph is a tree, the \num\QSAT[2] instance has value $n+1$; if the graph has a single cycle, it has value $2$; and if it has two or more cycles, it has value zero (\ie~it is unsatisfiable, or frustrated as a Hamiltonian).

The apparent reason for this is because a Haar-random state $\ket{\eta_{u,v}}$ is almost certainly entangled.
Following Refs.~\cite{Bravyi06,LMSS10,dBOE10}, if $h_{u,v}$ and $h_{v,w}$ project onto entangled states $\ket{\eta_{u,v}}$ and $\ket{\eta_{v,w}}$, a single-spin state on $u$ determines the feasible single-spin states at both $v$ and $w$ similarly to an instance of classical \XORSAT[2], in which the states of each interacting pair of bits strongly restrict each other. 
Typical instances of \QSAT[2] thus have effective long-range constraints between qubits within any connected component.
As a result, any graph which is dense enough to contain multiple cycles almost certainly gives rise to an overconstrained instance of \QSAT[2], corresponding to a frustrated Hamiltonian.
This is in contrast to \CNFSAT[2] formulae, which as instances of \QSAT[2] have constraints given by standard-basis vectors $\ket{\eta_{u,v}} = \ket{e_u}\ox\ket{e_v}$  for $e_u, e_v \in \{0,1\}$.
Such constraints on qubit-pairs $\{u,v\}$ and $\{v,w\}$ may fail to impose any constraints between the next-nearest neighbour qubits $u$ and $w$.
This is particularly important in the monotone special case of \num\SAT[2], which corresponds to \num\QSAT[2] instances in which $\ket{\eta_{u,v}} = \ket{00}_{u,v}$ for all edges $uv$ (corresponding to the constraint $x_u \vee x_v$ on boolean strings $x \in \{0,1\}^n$), which is itself \num\P-complete~\cite{Valiant79}.

\subsection{The typical difficulty of \num\QSAT[2] with product constraints}

To obtain instances of \num\QSAT[2] which resist solution by polynomial-time algorithms, there must be a substantial chance of obtaining tensor product constraints on each edge.
That this does not happen for Haar random constraints (a natural analogue to uniformly random constraints on pairs of bits) is a feature of quantum information theory, but does not shed much light on the range of difficulty of \num\QSAT[2].
We ask: which random graph families, and which distributions of constraints, yield difficult instances of \num\QSAT[2]?
Specifically, if only product constraints are involved, when is \num\QSAT[2] likely to be polynomial-time solvable?

We show, both for \ErdRen\ graphs and for bond-percolated rectangular lattices in two and three dimensions, that difficult instances of \num\QSAT[2] are rare if we select \iid\ product constraints from a distribution which differs substantially from monotone constraints.
In particular, on bond-percolated lattices, we expect the value of any \num\QSAT[2] instance to be efficiently solvable almost surely; and for \ErdRen\ graphs, the difficult-to-compute regime vanishes as the ``monotonicity'' of the constraint distribution decreases.

We may state our results more precisely, as follows.
We say that a property which holds \emph{asymptotically almost certainly (or surely)} is one which holds with probability $1 - O(1/{\poly(n)})$.
Following the usual terminology associated with the study of random graphs, we often omit the word ``asymptotically'' in connection with properties which hold almost surely/certainly: statements about discrete distributions which are ``almost'' certain or sure, are intended to be interpreted in the limit $n \to \infty$.
Considering (families of) Hamiltonians on $n$ qubits, we say that a system is \emph{highly disconnected} if its connected subsystems almost surely all have size $O(\log n)$; similarly, if it can almost surely be decomposed into subsystems of size $O(\log n)$ which are independent of one another (despite chains of intermediate interactions), we say that the system is \emph{highly decoupled}.
The following Lemma follows easily from the definitions of these terms: we discuss this in Section~\ref{sec:proofEasyHamiltonians}.
\begin{lemma}
	\label{lemma:easyHamiltonians}
  Instances of \num\QSAT[2] which are highly disconnected, frustrated, or highly decoupled are \emph{easy} (solvable in time $O(\poly n)$ on \eg~a deterministic Turing machine).
\end{lemma}
\noindent We consider constraint models interpolating between monotone \num\SAT[2] on one hand, and continuous probability density functions of product constraints on the other.
For ${f \ge 1}$, let $\vec q = (q_1, q_2, \ldots, q_f)$ be a distribution on $f$ distinct single-qubit states $\ket{\alpha_1}$, $\ket{\alpha_2}$, \ldots, $\ket{\alpha_f}$, used to generate constraints $\ket{\eta_{u,v}} = \ket{\alpha_u} \ox \ket{\alpha_v}$, where the factors are independently sampled from $\vec q$.
For example, $\vec q = (1,0,0,\ldots)$ for monotone \SAT[2], and $\vec q = (\tfrac{1}{2},\tfrac{1}{2},0,\ldots)$ for uniformly-random \SAT[2].
If $\mathbf q = ({1\!\!\;/\!\!\:f}\:\!, {1\!\!\;/\!\!\:f}\:\!, \ldots, {1\!\!\;/\!\!\:f}\:\!, 0, \ldots)$, we have $\| \mathbf q \|_2 = \| \mathbf q \|_\infty = {1\!\!\;/\!\!\:f}$, which approaches $0$ as $f \to \infty$; this limiting distribution is precisely that of single-qubit constraints chosen from the Haar measure.\footnote{%
	As $\vec q$ contains no information about the states $\ket{\alpha_j}$, we are glossing over how well-defined the limit $\vec q \to \vec 0$ is.
	We do not consider this here, but propose that $\left|\bracket{\alpha_j}{\alpha_k}\right| \le 1 - \Omega(1/f)$ for all $j \ne k$ should be sufficient to maintain a promise gap between the ground-state energy level and excited energy levels.}
Vector norms of $\vec q$ thus measure how monotone the random constraints typically are.
Let $Q_2 = 1 - \| \vec q \|_2^2$, and let $Q_\infty = 1 - \| \vec q \|_\infty$.
\begin{theorem}[\ErdRen\ models]
  For an \ErdRen\ graph on $n$ vertices with $m = \gamma n$ edges, instances of \num\QSAT[2] with $\gamma < \frac{1}{2}$ are almost certainly highly disconnected, and 
	instances with $\gamma > \frac{1}{2Q_2}$ are almost certainly frustrated; while if $2\gamma Q_\infty - \ln(2\gamma) > 1$, frustration-free instances are almost certainly highly decoupled.
\end{theorem}
\noindent --- thus, in the $\vec q \to \vec 0$ limit, a phase of typically ``difficult'' problems exists only for $m/n \sim \frac{1}{2}$.
\begin{theorem}[Bond-percolated lattice models]
	Let $d \in \{2,3\}$, and consider a $d$-dimensional square or cubic lattice on $n$ vertices: a segment of the rectangular grid $\Z \x \Z$ of dimensions $O(\sqrt n) \x O(\sqrt n)$, or of the cubic grid $\Z \x \Z \x \Z$ with dimensions $O(\sqrt[3]n) \x O(\sqrt[3]n) \x O(\sqrt[3]n)$, in which edges are present between nearest neighbours independently with some probability $p$.
	Let $p_c$ denote the critical percolation probability, at which there asymptotically almost surely exists a component of size $\Omega(n)$.
	For bond-percolated vertices with $m$ edges, if $Q_\infty$ is bounded away from $0$, there is a transition at $\frac{m}{dn} \in \Theta(n^{-1/7})$ from being almost certainly highly disconnected and frustration-free to being almost certainly frustrated.
	If we condition on frustration-free instances, we find instead that instances for which the percolation probability is subcritical (that is when $\frac{m}{dn} \le p_c$) are almost certainly highly disconnected, while instances for which $Q_\infty$ is greater than some constant $p_{\text{fin}} < 1$ (which depends on $d$) are almost certainly highly decoupled.
\end{theorem}
\noindent --- thus, a typical instance is almost surely solvable in polynomial-time even for $\vec q$ which deviates from monotonicity by only a finite amount.

The above results suggest that the only difficult instances of \num\QSAT[2] must be specially constructed to resemble monotone instances of \num\SAT[2].
Specifically: \textbf{(a)}~hard instances of \num\QSAT[2] are atypical, and \textbf{(b)}~the reason for this does not have to do with entangled constraints, but rather that an instance of \num\QSAT[2] is only likely to be difficult if its constraints are not very diverse and it is relatively sparsely constrained.

\smallskip\noindent
\textbf{Structure of this article.}
Section~\ref{sec:preliminaries} contains preliminary definitions and discussion, including techniques to infer long-range constraints, to count solutions to instances of \num\QSAT[2], 
and types of easily solved instances of \num\QSAT[2].
Section~\ref{sec:discreteProbabilisticModels} presents the conditions under which \num\QSAT[2] is easily solvable for instances whose interaction graphs are generated according to either the \ErdRen\ distribution or percolated rectangular/cubic lattice models.
In Section~\ref{sec:openProblems} we suggest some ways in which this work might be extended. 

\ifprelimanalysis
\section{Preliminaries}
\label{sec:preliminaries}

We consider \emph{simple graphs}, containing no parallel edges or single-vertex loops.
We denote the state-space of a generic qubit by $\cH_2 \cong \C^2$, and space of a particular qubit $u$ by $\cH_u$.
For the sake of brevity we occasionally neglect error terms which are decreasing in $n$: for instance, we write $f(n) \sim g(n)$ when $f(n) = g(n) \bigl[1 \pm o(1)\bigr]$ (which is an equivalence relation) and $f(n) \gtrsim g(n)$ when $f(n) > g(n) \bigl[1 \pm o(1)\bigr]$ (which is a quasi-order).

While \QSAT[2] allows for a broader range of constraints, in this article we consider only Hamiltonians $H = \sum h_{u,v}$, where $h_{u,v}$ is a rank-1 projector on $\C^2 \ox \C^2$ and the sum ranges over pairs of vertices $\{u,v\}$ which are adjacent in some graph (usually a typical graph from a given probability distribution on graphs).
It should be easy to see by extending the results below that instances of \QSAT[2] whose constraints correspond to projectors of rank $2$ or more will only increase the probability that the instance is efficiently solvable, by reason of the emergence of long-range constraints on the marginals of satisfying states.

For each rank-1 projector $h_{u,v}$, we consider the state $\ket{\eta_{u,v}} \in \cH_u \ox \cH_v$ such that
\begin{equation}
	h_{u,v} = \ket{\eta_{u,v}}\bra{\eta_{u,v}} \ox \idop_{V \setminus \ens{u,v}} \,. 
\end{equation}
For $H$ frustration-free, the operator $\bra{\eta_{u,v}}$ is a constraint on any ground-state $\ket{\psi}$ of $H$: for $\rho_{u,v}$ the density operator of $\ket{\psi}$ on $\ens{u,v}$, we have $\bra{\eta_{u,v}} \rho_{u,v} = 0$ by hypothesis.
Thus, as with the classical decision problem \SAT[2], we describe instances of \QSAT[2] by a list of local ``forbidden'' configurations $\bra{\eta_{u,v}} : \C^2 \to \C$  on pairs of qubits $u,v \in V$ (implicitly taking the tensor product with the identity on all other qubits) for a global state to avoid.

\subsection{Constraint induction}
\label{sec:constraintInduction}

Let $\ket{\Psi^-} \propto \ket{01} - \ket{10}$ be the singlet state.
Following Ref.~\cite{Bravyi06}, given constraints $\bra{\eta_{u,v}}, \bra{\eta_{v,w}}$ for $u \ne w$ which both act on a qubit $v \in V$, we may infer a further implicit constraint $\bra{\tilde \eta_{u,w}}$, such that $\bra{\tilde\eta_{u,w}} \rho_{u,w} = 0$ whenever both $\bra{\eta_{u,v}} \rho_{u,v} = 0$ and $\bra{\eta_{v,w}} \rho_{v,w} = 0$ hold:
\begin{equation}
	\label{eqn:constraintInduction}
  \bra{\tilde\eta_{u,w}}	\;\propto\,
		\Bigl[ \bra{\eta_{u,v}} \ox \bra{\eta_{v,w}} \Bigr] \Bigl[ \idop_u \ox \ket{\Psi^-} \ox \idop_w \Bigr].
\end{equation}
We may renormalise $\bra{\tilde\eta_{u,w}}$ so that $\bracket{\tilde\eta_{u,w}}{\tilde\eta_{u,w}} = 1$, provided that the operator is non-zero.
We may induce further implicit constraints recursively.
For two operators $\bra{\eta_{u,v}}$ and $\bra{\eta_{v,w}}$, we may write the operator obtained via Eqn.~\eqref{eqn:constraintInduction} by $\bra{\eta_{u,v}} \ast \bra{\eta_{v,w}}$.
It is easy to show that the binary operator ``$\ast$'' is associative, so that
\begin{equation}
	\label{eqn:constraintInductionAssoc}
  \bra{\eta_{u,v}} \ast \bra{\eta_{v,w}} \ast \bra{\eta_{w,x}} \;\propto\,
  \Bigl[ \bra{\eta_{u,v}} \ox \bra{\eta_{v,w}} \ox \bra{\eta_{w,x}} \Bigr] \Bigl[ \idop_u \ox \ket{\Psi^-} \ox \ket{\Psi^-} \ox \idop_x \Bigr],
\end{equation}
and so forth for longer chains, so that we may write
$
	\bra{\tilde\eta_{u,z}} = \bra{\eta_{u,v}} \ast \bra{\eta_{v,w}} \ast \cdots \ast \bra{\eta_{y,z}}
$ 
for an operator acting on $\ens{u,z}$ induced by a chain of constraints from the input instance of \QSAT[2].
This is similar, in the classical setting, to computing the transitive closure of the implication graph defined by Aspvall, Plass, and Tarjan~\cite{APT-1979}, in which case we may find multiple constraints between a pair of variables which tightly constrain their values.
Similarly, in the more general quantum setting, we may obtain multiple constraints $\bra{\smash{\eta^{(1)}_{u,v}}}, \bra{\smash{\eta^{(2)}_{u,v}}}, \ldots$ which may allow us to represent their joint state-space as a two-dimensional subspace $S \le \cH_u \ox \cH_v$, allowing us to reduce the number of qubits involved in the problem by a renormalisation step~\cite{dBOE10} without affecting the dimension of the space of satisfying states $\ket{\psi}$.

With respect to the operation ``$\ast$'' of induction of constraints, there are two significantly different constraint types: product constraints $\bra{\eta_{u,v}} = \bra{\alpha_u} \ox \bra{\beta_v}$, and entangled constraints which do not factor in this manner.
It is immediate that for $\bra{\tilde\eta_{u,w}} = \bra{\eta_{u,v}} \ast \bra{\eta_{v,w}}$, the constraint $\bra{\tilde\eta_{u,w}}$ is a product constraint if either $\bra{\eta_{u,v}}$ or $\bra{\eta_{v,w}}$ is; and that $\bra{\tilde\eta_{u,w}} = 0$ only if both $\bra{\eta_{u,v}} = \bra{\alpha_i} \ox \bra{\alpha_j}$ and $\bra{\eta_{v,w}} = \bra{\alpha_k} \ox \bra{\alpha_\ell}$ satisfy $\ket{\alpha_j} \propto \ket{\alpha_k}$.
When this occurs, the marginal state of $u$ cannot indirectly constrain the marginal on $w$, or vice-versa, through the interaction with $v$: by setting $v$ to the state $\ket{\bar\alpha_j}$ in the kernel of $\bra{\alpha_j}$, we extend any marginal on $\{u,w\}$ to one on $\{u,v,w\}$ which satisfies the constraints $\bra{\eta_{u,v}}$ and $\bra{\eta_{v,w}}$.

\subsection{Randomly generated instances of \num\QSAT[2]}

A ``random instance'' of \num\QSAT[2] is a sample from a probability distribution over instances of \num\QSAT[2], generally with a fixed number $n$ of qubits and $m$ of constraints.
We consider a generation process in which one first generates a random graph, either by selecting a fixed number $m$ of edges from the set of all possible pairs of edges (the \emph{\ErdRen\ graph model}), or by considering a subgraph of some lattice in which each lattice-edge is included with a probability $p$ such that the expected number of edges is $m$, associating a qubit to each vertex of the graph.
At each edge $uv$ in the random graph, we assign an operator $\bra{\eta_{u,v}}: \C^4 \to \C$ according to some probability distribution, 
representing two-body constraints on the qubits.

\label{sec:frustrationFreeGeneration}
We would like to also consider instances of \num\QSAT[2] which are guaranteed to have a non-zero value, corresponding to a distribution on two-body frustration-free Hamiltonians.
This requires a subtler random generation procedure.
For a model of random graphs (\eg~either an \ErdRen\ model or a percolated lattice model), we select a random order for the edge-set of the graph.
Adding these edges sequentially to graph, we assign a constraint to each, restricting the choice of constraint so that the resulting instance of \QSAT[2] is satisfiable.
In any continuous distribution (such as the Haar measure), any non-trivial restriction of the constraint model typically will be to a set of measure zero; the notion of restriction we intend is limit as $\epsilon \to 0$, of the Haar measure conditioned on being within an $\epsilon$-neighbourhood (in the Euclidean norm on $\C^4$) of the valid choices of constraint.
(For instance, if only a finite set of constraints avoid making the instance unsatisfiable, such a restriction yields the uniform distribution over those constraints.)
For the Haar measure, as well as for the product-constraint model of our article, there is always a choice of constraint for which the instance is satisfiable at each step: this is easy to show in the Haar random case by a minor extension of Ref.~\cite{LMSS10}, and can be established for the constraint model of this article without difficulty (see \eg~the beginning of Section~\ref{sec:discreteProbabilisticModels}).

\subsection{Remarks on the counting complexity of instances of \num\QSAT[2]}
\label{sec:countingTechniques}

Given a randomly generated instance of \num\QSAT[2], we ask: with what probability is it a ``difficult'' instance?
Our notion of ``difficulty'' is defined relative to some fixed algorithm $A$: a family of instances for which $A$ can successfully compute the answer in polynomial time are ``easy'', and families for which $A$ has no such upper bound are ``difficult''.
Such statements depend on the state of the art in combinatorics: an improved analysis of random graphs may show that some family of formerly ``difficult'' instances happen to be solvable by $A$ in polynomial time.
If one accepts standard complexity-theoretic assumptions such as $\P \ne \NP$, there are families of instances of \QSAT[2] which are inherently ``easy'' or ``difficult'' for any algorithm implemented \eg~on Turing machines.
The aim of this article is to establish bounds on the extent of any such ``difficult'' regime for certain distributions on \num\QSAT[2].

An instance of \QSAT[2] is \emph{monotone} if there is a state $\ket{\alpha_0} \in \C^2$ such that $\bra{\eta_{u,v}} = \bra{\alpha_0} \ox \bra{\alpha_0}$ for each $uv \in E(G)$.
This is equivalent to there being a local unitary operator $U$ such that $\bra{\eta_{u,v}} (U \ox U) = \bra{00}$ for all $uv \in E(G)$: the classical monotone instances of \num\SAT[2] are a special case in which we may take $U$ to be the identity.
As monotone \num\SAT[2] is \num\P-complete~\cite{Valiant79}, it follows that \num\QSAT[2] is at least \num\P-hard.
Ji, Wei, and Zeng~\cite{JWZ11} show that \num\QSAT[2] is also contained in \num\P, by a simple transformation of instances of \num\QSAT[2] which preserves the solution space and puts the interaction graph into a standard form.

Even monotone instances of \num\QSAT[2] may have structural properties which may render it ``easy''.
For instance, instances whose interaction graphs $G$ have bounded \emph{tree-width}~\cite{RS-1984} (see Ref.~\cite{Diestel-text} for an introductory reference) may be solved in $\poly(n)$ time,\footnote{%
	The approach here, for instances having tree-width at most $w > 0$, is essentially to use dynamic programming to count the partially-satisfying solutions for each of $2^w$ possible assignments (in some local basis) for each qubit indexed by a vertex in a tree-decomposition.
	A more complete description can be found in~\cite{FMR-2008}.%
}
albeit with a constant factor which grows exponentially with the tree-width~\cite{FMR-2008}.
This algorithm is useful in particular for tree graphs or connected graphs which have a single cycle, which respectively have tree-width $1$ and $2$.
Conversely, instances of \num\SAT[2] which are not monotone may still be ``difficult'': for a fixed graph $G$, if we assign a uniformly random clause to each $uv \in E(G)$, represented in the format of constraint operators for an instance of \num\QSAT[2] as one of the operators
$
	\bra{\eta_{u,v}} \in \bigl\{ \bra{00}, \bra{01}, \bra{10}, \bra{11} \bigr\}
$ 
then the non-trivial constraints arising between pairs of bits by the induction procedure of Eqn.~\eqref{eqn:constraintInduction} only extend over paths of expected length $O(1)$ in $G$.
Then only for sets of nodes where the constraints are relatively dense can there be a chance of giving rise to long-range constraints of order the size of a given connected component: this is necessary to impose enough structure to obtain an instance of \num\SAT[2] substantially different in complexity from a monotone instance on $n^{O(1)}$ variables.

\subsection{Three types of easily solved cases of \num\QSAT[2]}
\label{sec:proofEasyHamiltonians}

We now remark on the simple observations presented in Lemma~\ref{lemma:easyHamiltonians}: this will allow us to reduce the task of proving that instances of \num\QSAT[2] are easy, to showing that they fall into one of three structural classes of Hamiltonian --- \emph{frustrated}, \emph{highly disconnected}, or \emph{highly decoupled}, in the senses described preceding Lemma~\ref{lemma:easyHamiltonians}.

Following Chvatal and Reed~\cite{CR-1992} concerning phase transitions in the satisfiability of random instances of \CNFSAT[2], one may obtain results concerning random classical \num\SAT[2] on \ErdRen\ graphs with $n$ vertices and $m$ clauses.
Specifically, an instance of \SAT[2] with density $\frac{m}{n} > 1$ is almost certainly unsatisfiable, and so by definition has value zero as an instance of \num\SAT[2]; and this can be determined in polynomial time by detecting certain unsatisfiable substructures.
Similar remarks apply for \emph{frustrated} instances of \num\QSAT[2]: if one can efficiently determine that it is frustrated, this suffices to show that it has value zero.

As for easily solvable instances of \num\SAT[2] with positive values, if $\frac{m}{n} < \tfrac{1}{2}$, the underlying graph is almost certainly composed of components of size $O(\log n)$ having at most one cycle.
One can solve each such component in polynomial time using brute-force techniques (testing all possible assignments for each component); using dynamic programming and taking advantage of the existence of a tree decomposition for the component, one can even solve them in time linear in the component size (up to a logarithmic factor due to handling vertex labels for a graph of size $n$).
These represent a \emph{disconnected} regime in random \num\SAT[2]; and again, similar techniques apply for \num\QSAT[2] if we can establish that the components scale as $O(\log n)$, and/or have treewidth bounded by a constant as we have described above.
It then suffices to multiply the \num\QSAT[2] values for each component together: for random graph models (such as the ones we consider) where small components dominate, this may be done efficiently, \eg~using an algorithm which we describe in Appendix~\ref{apx:multiplyLongList}.

Finally, we may consider \emph{highly decoupled} instances, in which a subsystem which is contiguous nevertheless decomposes into independent subsystems of size $O(\log n)$.
These may arise in instances which have been constructed to be frustration-free, due to the proliferation of qubits whose states are ``fixed'' by their constraints.
When a qubit $x$ can only occupy a unique state in a satisfying state, we refer to this as the \emph{fixed state} of the qubit $x$ (which we denote $\ket{\bar\psi_x}$).
As we add constraints to a satisfiable instance of \QSAT[2], there are at least two ways in which an added constraint can increase the number of qubits with fixed states: 
\begin{romanum}
\item
	either by adding a constraint $\bra{\eta_{x,y}}$ between some qubit $x$, and a qubit $y$ which already has a fixed state such that $\bra{\eta_{x,y}} {\bigl(\idop_x \ox \ket{\smash{\bar\psi_y}}\bigr)} \ne \vec 0\herm$,
\item
	 or by adding a constraint which closes a chain of constraints starting and ending at $x$, which is only satisfiable by a single state $\ket{\bar\psi_x}$.
\end{romanum}
Any constraint $\bra{\eta_{x,y}}$ acting on a qubit $x$ with a fixed state will either be satisfied by $\ket{\bar\psi_x}$ regardless of the state of $y$, or will serve to fix the state of $y$.
Thus, interactions between qubits with fixed states with \emph{non}-fixed qubits will, by construction, fail to give rise to any long-range constraints between qubits without fixed states.
If there are enough qubits with fixed states, these may then effectively partition the set of \emph{non}-fixed qubits into independent subsystems; if these subsystems are of size $O(\log n)$, the system is then \emph{highly decoupled}.
Thus, to solve an instance of \num\QSAT[2], it also suffices to identify enough fixed qubits to partition the remainder into systems whose degeneracy may be efficiently computed.

Our result is to show how in two different random graph models, for random instances of \QSAT[2] with enough diversity in the constraints to differ substantially from monotone instances, there is (at most) a narrow range in which the density of constraints may give rise to instances which are neither highly disconnected, nor frustrated, nor highly decoupled almost surely.

\fi

\section{Discrete probabilistic models}
\label{sec:discreteProbabilisticModels}

We consider a constraint model of \emph{independent factor distributions}, in which constraints are product operators
$\bra{\alpha} \ox \bra{\beta}$ for some \iid\ single-qubit operators $\bra{\alpha}, \bra{\beta}: {\C^2 \to \C}$ distributed over some set of operators 
$
	\ens{\bra{\alpha_1}, \bra{\alpha_2}, \ldots, \bra{\alpha_f}}
$ 
for some $f \ge 1$, where $\bra{\alpha_j} \not\propto \bra{\alpha_k}$ for $j \ne k$.
Given an edge which represents a product constraint, the probability of obtaining $\bra{\eta_{u,v}} = \bra{\alpha_h}_u \ox \bra{\alpha_j}_v$ is given by $q_h q_j$, where 
$
		\vec q	\;=\;	(q_1,\, q_2,\, \ldots,\, q_f)
$ 
is a fixed probability distribution.
Throughout the following, we suppose that $1 > q_1 \ge q_2 \ge \cdots \ge q_f > 0$, so that there is some probability of obtaining non-monotone instances of \QSAT[2].

Independent factor distributions have convenient features for analysis.
Following Ref.~\cite{JWZ11}, the ground-state manifold for an instance of \QSAT[2] having only product constraints has a basis consisting of product states.
Furthermore, non-zero induced constraints ${\bra{\eta_{u,v}} \ast \bra{\eta_{v,w}}}$ range over the same two-qubit operators as the individual edge-constraints themselves (albeit with a different probability distribution than $\vec q \ox \vec q$).
As with Haar-random models, when we wish to consider only random \emph{frustration-free} Hamiltonians, we must specially select the constraints to meet that restriction.
We construct the random graph in the same manner as described in Section~\ref{sec:frustrationFreeGeneration}, this time restricting the choice of constraints according to the condition of not giving rise to a frustrated (\ie~an unsatisfiable) instance of \QSAT[2].
Frustration can only arise if both qubits on which the constraint are each restricted to some ``fixed'' state to satisfy the earlier constraints placed on it: a ``non-frustrating'' choice of constraint can then be made simply by having it be satisfied by one of the two fixed states.

We may consider how likely long-range constraints (as described in Section~\ref{sec:constraintInduction}) are for such a constraint model.
Let $x_0, x_\ell \in V(G)$ be two vertices connected by a path $P = x_0 x_1 \cdots x_\ell$ in the interaction graph of a random instance of \num\QSAT[2].
We may consider what constraints may exist on the joint state of $x_0$ and $x_\ell$ by virtue of the inducted constraint
$
		\cC_P	=	\bra{\eta_{x_0,x_1}} \ast \bra{\eta_{x_1,x_2}} \ast \cdots \ast \bra{\eta_{x_{\ell\text{--}1},x_\ell}}
$. 
One may show by induction that $\cC_P$ is non-zero if and only if $\bra{\eta_{x_{h\text{--}1},x_h}} \ast \bra{\eta_{x_h,x_{h+1}}} \ne 0$ for each index $0 < h < \ell$ of internal vertices of the path.
For each such $h$, we have $\bra{\eta_{x_{h\text{--}1},x_h}} \ast \bra{\eta_{x_h,x_{h+1}}} = 0$ if and only if 
$
  \bra{\eta_{x_{h\text{--}1}, x_h}} = \bra{\alpha_i}\ox\bra{\alpha_j}
$ and $
  \bra{\eta_{x_{h\text{--}1}, x_h}} = \bra{\alpha_k}\ox\bra{\alpha_\ell}
$ 
for some $j \ne k$.
Because the right-factor of $\bra{\eta_{x_{h\text{--}1},x_h}}$ and the left-factor of $\bra{\eta_{x_h,x_{h+1}}}$ are independently distributed, this occurs with probability
\begin{equation}
	Q_2 \;:=\; 1 - \| \vec q \|_2^2	\;=\;	\sum_{j=1}^f q_j(1-q_j)	\;\le\; 1 - \frac{1}{f}\,,
\end{equation}
with equality if and only if $\vec q$ is uniform.
Note that $Q_2 > 0$, where the lower bound is the infimum as $\vec q \to (1,0,0,\ldots)$. 
As the probabilities of having identical factors at each vertex are independent, we then have
\begin{equation}
	\label{eqn:chainConstraintNonzeroIndptFactors}
	\Pr\Bigl[\cC_P \ne 0\Bigr]	\;=\;	\prod_{h=1}^{\ell-1} \left(1 - \| \vec q \|_2^2 \right)	\;=\;	Q_2^{\ell-1}\,.
\end{equation}
Thus, $\cC_P$ is non-zero and proportional to $\bra{\alpha_h} \ox \bra{\alpha_j}$ with probability $q_h q_j Q_2^{\ell-1}$ for each ${1 \le h, j \le f}$, and equal to zero with probability $1 - Q_2^{\ell-1}$.
Because the long-range constraints which involve a particular vertex as a \emph{mid-point} are not independent of one another, it may be useful in some cases to bound this probability from below by $Q_\infty^{\ell - 1}$, where $Q_\infty = 1 - \| \vec q \|_\infty$, where $\| \vec q \|_\infty = q_1$ is an upper bound on the probability that the single-qubit operators $\bra{\alpha_j}, \bra{\alpha_k}$ with which two different constraints act on $x$ are the same.

\subsection{\ErdRen\ interaction graphs}
\label{sec:discreteProbabilisticErdRen}

The attenuation of the probability of long-range constraints described in Eqn.~\eqref{eqn:chainConstraintNonzeroIndptFactors} is similar to what occurs in uniformly random \SAT[2].
For \ErdRen\ interaction graphs on $n$ vertices and $m$ edges --- a distribution on labelled graphs which may be sampled by listing each of the $\binom{n}{2}$ potential edges in a random order, and selecting the first $m$ edges for inclusion --- this motivates an analysis which follows closely to that of Chvatal and Reed~\cite{CR-1992}, adapting it for counting problems and to involve more general constraint distributions.
We show that, except for a ``difficult phase'' in the regime $\frac{1}{2} \le \frac{m}{n} \le \frac{1}{2Q_2}$, a random instance of \num\QSAT[2] is almost certainly either highly disconnected or frustrated, according to whether $\frac{m}{n}$ is below or above the boundaries of the difficult phase.
In particular, the difficult phase shrinks to a band of zero width at $\frac{m}{n} \sim \frac{1}{2}$ as $Q_2 \to 1$.
In the special case of frustration-free instances, this band expands to $\frac{1}{2} \le \frac{m}{n} \le \frac{1}{2 Q_\infty} (1 + \delta)$ for some small $\delta$ which vanishes as $Q_\infty \to 1$; this band also converges to $\frac{m}{n} \sim \frac{1}{2}$ as $Q_\infty \to 1$.
Thus in the ``completely non-monotonic'' limit $\vec q \to \vec 0$, \num\QSAT[2] is always easy; and there is a substantial band of instances which may be difficult to solve only if the constraint distribution shows a corresponding bias towards a small, finite number of constraints.

\subsubsection{The highly disconnected phase in \ErdRen\ models}
\label{sec:highlyDisconErdRen}

Whether or not we restrict to frustration-free instances of \QSAT[2], the existence of a highly disconnected regime in instances of \QSAT[2] on \ErdRen\ graphs $G$ follows directly from the random graph model itself.
For $\frac{m}{n} < \frac{1}{2}$, almost certainly $G$ contains only components of size $O(\log n)$, and almost certainly contains no components having more than one cycle~\cite{ErdRen60}.
Any instance of \QSAT[2] on such a graph will thus be highly disconnected, regardless of the constraint distribution.
For our results on \ErdRen\ models, it thus suffices to establish upper bounds on the extent of any difficult phase.

\subsubsection{The frustrated phase in unconditional \ErdRen\ models}
\label{sec:critUnconditionalErdRen}

\begin{figure}
		\begin{minipage}{0.32\textwidth}
		\caption{%
			Example of a ``figure eight'' graph on ${2\ell-1}$ vertices, for $\ell = 8$.
			By Eqn.~\eqref{eqn:conditionalFrustrBicycleIndptFactors}, the probability of such a graph describing a frustrated figure-eight subsystem scales as $O\bigl(Q_2^{2\ell}\bigr)$.
		}
		\label{fig:figeight}
		\end{minipage}
		~\hfill
		\begin{minipage}{0.55\textwidth}
	\begin{tikzpicture}[scale=0.78,
		every node/.style={circle,fill=black,outer sep=-0.5ex}
		]

		\def\r{2}
	\foreach \o/\j in {-\r/b,\r/a} {
		\foreach \l/\d in {0/0,1/45,2/90,3/135,4/180,5/225,6/270,7/315} {
		\coordinate (\j\l) at ($(-\o,0) + (\d:\r)$);
	}

	\foreach \a/\b in {0/1,1/2,2/3,3/4,4/5,5/6,6/7,7/0} {
		\draw [line width=2pt, blue!85!white] (\j\a) --	(\j\b);
	}
}

	\foreach \o/\j in {-\r/b,\r/a} {
		\foreach \l/\d in {0/0,1/45,2/90,3/135,4/180,5/225,6/270,7/315} {
		\node (\j\l) at ($(-\o,0) + (\d:\r)$) {};
	}
	}

	\node at (a0) [label=left:${x_0 = x_\ell\,}$] {};
	\node at (a0) [label=right:${\, = x_{2\ell}}$] {};
	\node at (a1) [label=225:$x_1$] {};
	\node at (a2) [label=south:$\cdots$] {};
	\node at (a6) [label=north:$\ldots$] {};
	\node at (a7) [label=135:$x_{\ell-1}$] {};
	\node at (b3) [label=305:$x_{\ell+1}$] {};
	\node at (b2) [label=south:$\cdots$] {};
	\node at (b6) [label=north:$\ldots$] {};
	\node at (b5) [label=45:$x_{2\ell-1}$] {};
	\node at (a4) [outer sep=1.5em, label=right:{\Large$X_1$}] {};
	\node at (b0) [outer sep=1.5em, label=left:{\Large$X_2$}] {};
	\end{tikzpicture}
		\end{minipage}
		\\[2ex]
	\end{figure}
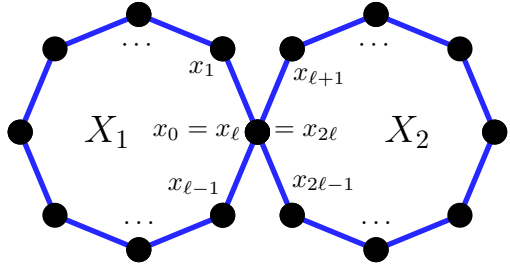

For a random graph with $m \in \Omega(n)$ edges, we adapt the analysis of Chvatal and Reed~\cite[Theorem~4]{CR-1992} to consider the probability that the giant component $\Gamma$ contains a ``frustrated figure eight'' (corresponding to a ``snake'' in Ref.~\cite{CR-1992}): a subsystem $X$ such that
\begin{enumerate}[label=(\textit{\roman*})]
\item
	Its interaction graph contains a \emph{figure eight graph}, which we define as a pair of cycles $X_1 = x_0 x_1 \cdots x_{\ell-1} x_\ell$ and $X_2 = x_\ell x_{\ell+1} \cdots x_{2\ell-1} x_{2\ell}$ of the same length, where $x_0 = x_\ell = x_{2\ell}$, and where $X_1$ and $X_2$ intersect only at the vertex $x_0 = x_\ell$.
	(See Fig.~\ref{fig:figeight} for an example.) 
	There may be additional edges connecting vertex-pairs $x_j x_k$ (though these will typically be unlikely), and $X = X_1 \union X_2$ may be connected to other vertices.\\[-2ex]

\item
	For each ${0 \le j < 2\ell}$, the constraints $\bra{\eta_{x_j\!\!\:,\:\!x_{j+1}}} = \bra{\beta_j} \ox \bra{\gamma_j}$ satisfy $\bra{\gamma_j} \ne \bra{\beta_{j+1}}$.
	\\[-2ex]
	
\item
	We have $\bigl\{\bra{\beta_0}, \bra{\gamma_{\ell-1}} \bigr\} \cap \bigl\{\bra{\beta_{\ell}}, \bra{\gamma_{2\ell-1}}\bigr\} = \varnothing$, so that the constraints imposed by $X_1$ and $X_2$ on their common spin $x_0$ are not simultaneously satisfiable.
\end{enumerate}
The cycles $X_1$ and $X_2$ are either ``alternating loops'' or ``quasi-alternating loops'' in the terminology of Ref.~\cite{JWZ11}, and impose constraints on $x_0$ is which cannot be simultaneously satisfied. 
Thus a frustrated figure eight is unsatisfiable by construction.
We consider the probability of a \emph{large} frustrated figure-eight arising in a random instance of \QSAT[2] with constraints given by an independent factor distribution, which in particular implies that it is part of the largest contiguous subsystem of the Hamiltonian.

In a system with a figure-eight subgraph, the probability of $\bra{\gamma_{j-1}} \ne \bra{\beta_j}$ is simply $Q_2$ for each of the $2\ell - 2$ sites $x_j$ of the two cycles, excluding the shared vertex $x_0 = x_\ell = x_{2\ell}$\,.
The conditions at the node $x_\ell$, where we require $\bra{\beta_0} = \bra{\gamma_{\ell-1}} \ne \bra{\beta_{\ell}} = \bra{\gamma_{2\ell-1}}$, occur with a probability $Q_{\text{crux}}$ which also depends only on $\vec q$.
(By~a routine calculation, one may show that
\begin{equation}
\begin{aligned}[b]
		Q_{\text{crux}}
  &=
		\;\sum_{h} q_h \biggl( \biggl[ q_h \!\sum_{j,k\ne h} \!\!q_j q_k  \biggr] + \biggl[ \,\sum_{i\ne h} \,q_i \!\!\!\!\sum_{j,k \notin \{h,i\}} \!\!\! q_j q_k  \biggr]\biggr)
	\\[1ex]&=
		 1 - 4 \| \vec q \|_2^2 + 2 \| \vec q \|_2^4 + 4 \| \vec q \|_3^3 - 3 \| \vec q \|_4^4\;.
\end{aligned}
\end{equation}
Then $Q_{\text{crux}} \to 1$ as $\vec q \to \vec 0$, and $Q_{\text{crux}} \in \Theta(1)$ for $\| \vec q \|_\infty$ bounded away from $1$.)
Given a fixed figure-eight graph $X$ on $2\ell - 1$ vertices, the probability that it gives rise to a \emph{frustrated} figure-eight system is then
\begin{equation}
\begin{aligned}[b]
	\label{eqn:conditionalFrustrBicycleIndptFactors}
		\Pr\Bigl[X \text{ a frustrated subsystem} \Bigr]\;=&\; Q_2^{2\ell-2} Q_{\text{crux}}.
\end{aligned} 
\end{equation}
Let $m = \gamma n$ for some constant $\gamma > 0$.
Using a second moment probabilistic argument, adapting the proof of Ref.~\cite[Theorem~4]{CR-1992}, we show that 
the largest contiguous subsystem almost certainly contains a frustrated figure eight so long as $\gamma > \frac{1}{2Q_2}$.

Let $\varphi_\ell$ denote the number of frustrated figure eight subsystems in $G$ on $2\ell-1$ vertices.
The mean $\bE(\varphi_{\ell})$ over all random graphs on $n$ vertices and $m$ edges can be evaluated by considering all sets $S$ of $2\ell-1$ vertices, and summing the probability of $S$ being such figure eight subsystem for all such subsets.
We will make use of the equality
\begin{equation}
	\label{eqn:permutationApprox}
  \frac{n!}{(n-t)!} \;\sim\;	n^t \exp\bigl(-\alpha(n,t) \bigr),
  \qquad
  \text{where }~
		\alpha(n,t)	:=	t + (n-t+\tfrac{1}{2}) \ln\!\left(\!1 - \frac{t}{n}\right)	
\end{equation}
which holds for $t \in o(n)$,
\footnote{%
	This may be easily recovered using Stirling's approximation.
}
ignoring a relative error term of $O(\tfrac{1}{n})$ using the notation defined at the beginning of Section~\ref{sec:preliminaries}.
By considering \parit{i}~the number of ways that we may choose the common vertex, \parit{ii}~the number of distinguishable ways that we may construct two cycles on $\ell$ vertices (built in either order) which incorporate the common node, \parit{iii}~the number of ways of allocating the remaining edges after having built $X$, and \parit{iv}~the probability that $X$ is a frustrated figure eight given that it is present in the random graph, we may obtain
\begin{equation}
	\label{eqn:meanFrustrFigureEights}
	\mspace{-18mu}
	\mbox{\small$
	\begin{aligned}[b]
			\bE(\varphi_\ell)
	\,&=\,
			Q_2^{2\ell-2} Q_{\text{crux}} \, \cdot \frac{n}{2}\left[ \tfrac{1}{2} \binom{n-1}{\ell-1} (\ell-1)! \right]
			\left[ \tfrac{1}{2} \binom{n-\ell}{\ell-1} (\ell-1)! \right]
			\left[ \frac{\displaystyle\binom{\binom{n}{2} - 2\ell + 1}{m-2\ell+1}}{\displaystyle\binom{\binom{n}{2}}{m}} \right]
	\\&\sim\,
		\frac{Q_{\text{crux}}}{8Q_2} 
		\left( \frac{2Q_2m}{n} \right)^{\!2\ell-1} 
		\frac{\exp\Bigl(-\alpha\bigl(n,2\ell-1\bigr) + \alpha\bigl(\tbinom{n}{2},2\ell-1\bigr) \Bigr)}{\exp\Bigl( \alpha\bigl(m,2\ell-1\bigr) \Bigr)}
		\;.
	\end{aligned}
	$}
	\mspace{-30mu}
\end{equation}
For $\ell \in o(n^{1/2})$, we have $\alpha(n,2\ell-1) \in o(1)$; then we can easily show that $\varphi_\ell > 0$ with non-zero probability, provided that $m = {\tfrac{1}{2Q_2}\bigl(1 + \Omega(\tfrac{1}{\ell})\bigr) n}$.

Next, we show that $\varphi_\ell$ almost surely doesn't differ substantially from its mean. 
Define a random variable $\varphi_X \in \ens{0,1}$ such that $\varphi_X = 1$ for instances of \QSAT[2] whose constraint subgraph contains a frustrated figure-eight on a given subgraph $X$.
We compare $\bE(\varphi_X)^2$ against $\bE(\varphi_X \varphi_Y)$, where $X = x_0 x_1 \cdots x_{2\ell-1} x_0$ and $Y = y_0 y_1 \cdots y_{2\ell-1} y_0$ are both figure-eight graphs on $2\ell - 1$ vertices, but which may have vertices and edges in common.
By definition, we have $\Var(\varphi_\ell) = \bE(\varphi_\ell^2) - \bE(\varphi_\ell)^2$.
We have
\begin{align}
  \bE(\varphi_\ell) \;&=\; \sum_X \Pr\bigl[ \varphi_X = 1\bigr]	,
&
	\bE(\varphi_\ell^2) \;&=\;	\sum_{X,Y} \Pr\bigl[ \varphi_X \varphi_Y = 1 \bigr],
\end{align}
where we sum over all possible figure-eight subgraphs $X,Y$ on $2\ell - 1$ vertices selected from $n$ vertices.
We show that $\bE(\varphi_\ell^2) \sim \bE(\varphi_\ell)^2$, which implies that $\Var(\varphi_\ell) \in o\bigl(\bE(\varphi_\ell)^2\bigr)$.

Consider the probability that a given subgraph $g$ on $t$ edges occurs as a subgraph of $G$.
Accounting for how we can distribute $t$ edges among the first $m$ elements of a random sequence of edges, we have
\begin{equation}
\begin{aligned}[b]
			f(t)
		\;:=\;
			\Pr\bigl[g \subset G\bigr]
		\;&=\;
			\binom{m}{t} \;\!t! \left[ \frac{\bigl(\binom{n}{2}-t\bigr)!}{\binom{n}{2}!}\right]
		\\&\sim\; 
			\left(\frac{2 \gamma}{n}\right)^{\!t} \exp\Bigl(\alpha\bigl(\tbinom{n}{2},t\bigr)-\alpha(\gamma n,t)\Bigr)
	\;.
\end{aligned}
\end{equation}
We suppose that $\ell \in o(n^{1/2})$, so that
$
			f(2\ell + \delta(\ell)) \;\sim\;
			(2 \gamma /n )^{\!2\ell + \delta(\ell)} 
$ 
for $\delta(\ell) \in \pm O(\ell)$, again using $\e^{\alpha(N,t)} \sim 1$ for $t \in o(N^{1/2})$.
For figure-eight subgraphs $X,Y$ on $2\ell-1$ vertices each, write $\Phi(X,Y) 
:= {\Pr\bigl[ \varphi_X \varphi_Y = 1 \,\big|\, X \union Y \subset G \bigr]}$ for the probability of the frustration conditions on $X \cup Y$.
Then if $\bigl\lvert E(X) \cap E(Y) \bigr\rvert = i$, 
\begin{equation}
		\Pr\Bigl[\varphi_X \varphi_Y = 1 \Bigr] 
	\;=\; \Pr\Bigl[ X \union Y \subset G \Bigr] \Phi(X,Y)
	\;=\;
		f(4\ell-i) \Phi(X,Y).
\end{equation}
For $X$ fixed, define $\Phi_i(X)$ to be the sum of ${\Pr\bigl[\varphi_X \varphi_Y = 1 \bigr]}$ over all figure-eight subgraphs $Y$ of the same size, for which $\left\lvert E(X) \cap E(Y) \right\rvert = i$ as above (\ie~the probability of obtaining two frustrated figure-eight subsystems which intersect in this way, one of which is $X$).
The probability of having any pair of isomorphic frustrated figure eight subgraphs, of which one is $X$, is then given by $\Phi(X) := \sum_i \Phi_i(X)$.

We may show that for a fixed $X$, the contribution of $\Phi_0(X)$ is the only significant contribution to $\Phi(X)$.
Note that if none of the edges of $X$ and $Y$ overlap, the frustration conditions for $X$ and for $Y$ are completely independent, even if $X$ and $Y$ share vertices: that is, $\Phi(X,Y) = \bigl[ Q_2^{2\ell-2} Q_{\text{crux}} \bigr]^2$ in this case.
We can then upper bound $\Phi_0(X)$ roughly by removing the restriction on $Y$ that $X \wedge Y$ have no edges.
\begin{subequations}
Let $F_{2\ell-1}$ denote the number of possible frustrated figure eight graphs on $2\ell - 1$ vertices selected from $n$ vertices: then 
\begin{equation}
\begin{aligned}[b]
		\Phi_0(X)
	\;&<\;
		\sum_Y f(4\ell) \Phi(X,Y) 
	\;\sim\;
		F_{2\ell-1} \left(2 \gamma \:\!\big/\;\!\! n \right)^{\!4\ell} Q_2^{4\ell-4} Q_{\text{crux}}^2 .
\end{aligned}
\end{equation}
For all other $0 < i \le 2\ell$, we consider the number $N(i,j)$ of figure-eight subgraphs $Y$ on $2\ell-1$ vertices, for which ${X\wedge Y}$ has $i$ edges and $j$ vertices, and consider an upper bound $\Phi(i,j)$ for the frustration probabilities $\Phi(X,Y)$ for all such subgraphs $Y$.
Then we have
\begin{equation}
		\Phi_i(X)
	\;\le\;
		\sum_j N(i,j) f(4\ell - i) \Phi(i,j)
\end{equation}
for $i > 0$.
We bound the parameters $\Phi(i,j)$ and $N(i,j)$ by considering bounds on the frustration conditions holding at each site in ${X \cup Y}$, and by considering how the number of components in ${X \wedge Y}$ affects both $N(i,j)$ and the probability of all the local frustration conditions holding.
\end{subequations}

\smallskip
\noindent
\textbf{Local frustration conditions.}
If $X$ and $Y$ intersect at all, the probabilities of the frustration conditions holding for any shared vertex only differs from what it would be independently for $X$ and for $Y$ if they also share edges.
For instance, if $x_j = y_k$ for $j,k \notin \ens{0,\ell,2\ell}$, and $e_{x,j}, e_{x,j+1} \notin E(Y)$, then the frustration conditions for $X$ and for $Y$ at $x_j$ are independent of one another and obtain with probability $Q_2^2$, as if $x_j$ and $y_k$ were actually distinct vertices.
Similarly, if $x_j = y_k$ for $j,k \notin \ens{0,\ell,2\ell}$, and $e_{x,j}, e_{x,j+1} \in E(Y)$, then the frustration conditions are identical and they obtain with probability $Q_2$.
The most interesting cases are for the ``crux'' vertices $x_\ell$ and $y_\ell$, and for the ``junction'' vertices of degree $3$ in $X \union Y$ arising from $x_j = y_k$ for $j,k \notin \ens{0,\ell,2\ell}$.
\begin{itemize}
\item 
	Vertices in $X \union Y$ of degree $3$ correspond to vertices $x_j = y_k$ for $j,k \notin \ens{0,\ell,2\ell}$, where one of the edges $e_{x,j}$ or $e_{x,j+1}$ is equal to one of the edges $e_{y,k}$ or $e_{y,k+1}$.
	To satisfy the frustration conditions, the common edge of $X$ and $Y$ which is adjacent to $x_j$ must act on $x_j$ differently from the remaining two edges, but the other two edges may act on $x_j$ in either distinct or identical ways to each other.
	Routine calculation shows that the probability of this occurring is $Q_{\text{junct}} := \| \vec q \|_2^2 - \|\vec q \|_3^3$.
\item
	The probability that the frustration conditions for $X$ holds at $x_\ell$, when $x_\ell = y_k$ for some $0 < k < 2\ell$, may be somewhat complicated if some of the edges of $Y$ incident to $y_k$ overlap some of the edges $\ens{e_{x,1}, e_{x,\ell}, e_{x,\ell+1}, e_{x,2\ell}}$ incident to $x_\ell$.
	Similar remarks apply to the other crux vertex $y_\ell$.
	As there are at most two crux vertices in $X \union Y$, we may ultimately subsume the probability that these conditions hold at $x_\ell$ or at $y_\ell$ as a constant factor, and simply bound the probability from above by $1$.
\end{itemize}

\smallskip
\noindent
\textbf{Vertex types and simultaneous frustration.}
The probability of $X$ and $Y$ both being frustrated depends on the number of junction vertices, crux vertices, and other vertices in $X \cup Y$, which are closely related to the number of components.
Extending the observation made with respect to the probability of frustration conditions holding at the crux vertices, we adopt an approach of avoiding case analysis, by sweeping various scalar factors under the rug when they depend only on a constant number of vertices.
To do so, we define a scalar factor $c$ (which we do not explicitly calculate) to bound from above any contributions by constant factors in the various cases.

In most cases, the components of $X \wedge Y$ (if it is non-empty) will consist of paths, and possibly one non-path tree component in the case that $x_\ell = y_\ell$ (with at least three of the edges of $X$ and $Y$ overlapping at that vertex).
In rare cases, $X \wedge Y$ may have a component which contains an entire cycle, or indeed two cycles if $X = Y$.
In the typical case where $X \wedge Y$ is cycle-free, 
the number of components will be the difference $j - i$;
Otherwise, $X \wedge Y$ has one or two cycles, so that it has $j-i+1$ or $j-i+2$ components.
In any case, the number of components is $j - i + O(1)$.
We may then make the following remarks concerning vertices of different types:
\begin{itemize}
\item 
	As we note above, $X \cup Y$ has at most $O(1)$ distinct crux vertices, for which frustration conditions occur with constant probability regardless of the number of edges of $X$ and $Y$ which overlap at those vertices.
\item
	The number of junction vertices is minimized when each component of $X \wedge Y$ is a path segment, with each component having two junction vertices at its endpoints; the largest number of junction vertices a component may have is four, in the case that the the two crux vertices coincide so that one component of $X \wedge Y$ has four leaf nodes. (Three junction nodes are possible as well if the two crux nodes coincide, but where only three of the edges of $X$ and of $Y$ coincide.)
	Thus the number of junction vertices is $2(j - i) + O(1)$ in all cases.

\item
	The frustration conditions elsewhere are governed by edge-pairs meeting at some vertex, where either both edges are common to $X$ and $Y$ or both belong to one figure-eight graph $X$ and $Y$ (the same one), but not to both.
	Considering the edges $x_0 x_1$, $x_1 x_2$, \etc\ in sequence and pairing each with the one that follows it, we may count these edge-pairs by considering those edges $x_j x_{j+1}$ for which $x_{j+1}$ is not a junction or crux vertex.
	The number of edges in $X$ which meet at non-junction, non-crux vertices is $2\ell - 2(j-i) - O(1)$, and similarly for $Y$; and the number of such edges in $X \wedge Y$ is $i - 2(j-i) - O(1)$, yielding a total of $4\ell - 2(j-i) - i \pm O(1)$.
\end{itemize}
\begin{subequations}
Thus for $0 < i < 2\ell$ we have
\begin{equation}
\label{eqn:jointFrustrationBound}
	\Phi(X,Y)
\;\le\;
	c \;\! Q_2^{4\ell-i} \left( \frac{Q_{\text{junct}}}{Q_2} \right)^{\!2(j-i)}
\end{equation}
for some constant $c$ depending only on the probability distribution $\vec q$ of constraint probabilities.
For $i = 2\ell$, we have $X = Y$ and $j = 2\ell - 1$: then following Eqn.~\eqref{eqn:conditionalFrustrBicycleIndptFactors} we may explicitly evaluate
\begin{equation}
\mspace{-10mu}
\begin{aligned}[b]
	\Phi(X,X) \,&=\, \Pr\bigl[ \varphi_{\!\!\:X} = 1 \,\big|\, X \subset G \bigr] \,=\, Q_{\text{crux}} Q_2^{2\ell-2}
	.
\end{aligned}
\mspace{-10mu}
\end{equation}
\end{subequations}

\smallskip
\noindent
\textbf{Ways to overlap at $i$ edges.}
Following the analysis of Ref.~\cite{CR-1992}, we may bound $N(i,j)$ by considering upper bounds on \parit{i}~the number of ways a fixed shape for the graph $X \wedge Y$ could be mapped injectively into $X$ and into $Y$, \parit{ii}~the number of ways that the components of $X \wedge Y$ could be arranged into the vertex-order of $Y$, and \parit{iii}~the number vertices which may belong to $Y \setminus (X \wedge Y)$.
\begin{subequations}
The number of subgraphs $Y$ such that $X \wedge Y$ has $i$ edges and $j$ vertices can then be bounded by
\begin{equation}
	\begin{aligned}[b]
				N_\ell(i,j)	\;&<\;	4 \binom{2\ell+2}{2j - 2i + 2}^{\!2} \ell \;\! (j-i)! \;\! 2^{j-i} \;\! n^{2\ell-j-1}
		\\[1ex]&\le\;
			4\ell (2\ell+2)^{4(j - i) + 4} \;\! 2^{j-i} \;\! n^{2\ell-j-1}
	\end{aligned}
\end{equation}
in the case $0 < i < \ell$, and
\begin{equation}
	\begin{aligned}[b]
				N_\ell(i,j)	\;&<\;	4 \binom{2\ell+2}{2j - 2i + 2}^{\!2} \ell \;\! (j-i)! \;\! 2^{j-i+1} \;\! n^{2\ell-j-1}
		\\[1ex]&\le\;
			8\ell (2\ell+2)^{4(j - i) + 4} \;\! 2^{j-i} \;\! n^{2\ell-j-1}
	\end{aligned}
\end{equation}
for $0 < i < 2\ell$ more generally. 
If for the sake of brevity we define $\Lambda = {2(2\ell+2)^4/n}$, we then have
\begin{equation}
	N_\ell(i,j) \le \begin{cases}
										2 \ell \Lambda^{j-i+1} n^{2\ell-i} & \text{if $0 < i < \ell$}, \\
	                  4 \ell \Lambda^{j-i+1} n^{2\ell-i} & \text{if $\ell \le i < 2\ell$}.
	                \end{cases}
\end{equation}
\end{subequations}
Again, we have $X = Y$ if $i = 2\ell$, so that $N_\ell(2\ell,j) = 1$.

\bigskip
Suppose that $\ell \in o(n^{1/4})$, so that $\Lambda \in o(1)$.
We may then use the above remarks to bound $\Phi_i(X)$ for $i > 0$. 
For ${0 < i < \ell}$, the graph $X \wedge Y$ has no cycles, so that ${i+1 \le j \le 2\ell-1}$; 
\begin{subequations}
we may then bound
\begin{align}
\mspace{-6mu}
		\Phi_i(X)
	\;&\le
		\sum_{j = i+1}^{2\ell-1} 	\!
			N(i,j) \Phi(i,j) f(4\ell-i) 
	\notag\\&<
		\sum_{j = i+1}^{2\ell-1} 	
			\Biggl[ 2\ell \Lambda^{j-i+1} n^{2\ell-i} \Biggr] 
			\left[
				c Q_2^{4\ell-i} \left( \frac{Q_{\text{junct}}}{Q_2} \right)^{\!\!2(j-i)} \right] \left(2 \gamma \:\!\big/\;\!\! n \right)^{\!4\ell-i}
	\notag\\&=\,
		2c\ell \Lambda n^{2\ell-i} Q_2^{4\ell-i} \left(2 \gamma \:\!\big/\;\!\! n \right)^{\!4\ell-i} 
		\sum_{j = i+1}^{2\ell-1} 
				\left( \frac{\Lambda Q_{\text{junct}}^2}{Q_2^2} \right)^{\!\!j-i}
	\notag\\[1ex]&<\,
		2c\ell \Lambda n^{-2\ell}
		\left(2 \gamma Q_2 \right)^{\!4\ell-i} 
		\left( \frac{\Lambda   Q_{\text{junct}}^2}{Q_2^2} \right) 
		\left( \frac{1}{1 - {\Lambda Q_{\text{junct}}^2 Q_2^{-2}}} \right)
	\notag\\[1ex]&\sim\,
		\left(\frac{2cQ_{\text{junct}}^2}{Q_2^2}\right)
		\ell \Lambda^2 n^{-2\ell}
		\left(2 \gamma Q_2 \right)^{\!4\ell-i}
	.
\end{align}
For ${\ell < i < 2\ell}$, we may only bound ${i \le j \le 2\ell-1}$, and for $i = 2\ell$ we have ${j = 2\ell-1} = {i-1}$; we may then obtain similar bounds
\begin{align}
	\mspace{-12mu}
		\Phi_i(X)
	\;&\lesssim\;
		4c
		\ell \Lambda n^{-2\ell}
		\left(2 \gamma Q_2 \right)^{\!4\ell-i}
	&&\text{for $\ell \le i < 2\ell$,}\!
\\
	\mspace{-12mu}
		\Phi_{2\ell}(X)
	\;&\sim\;
		Q_{\text{crux}} Q_2^{-2} n^{-2\ell}
		\left(2 \gamma Q_2 \right)^{\!2\ell}
	\!\!\!
	&&\text{for $i = 2\ell$.}\!
\end{align}
\end{subequations}
Expanding the formulas for $\Phi_i(X)$ for $i > 0$ and eliding the constant factors, we may obtain
\begin{equation}
\mspace{-19mu}
\begin{aligned}[b]
		\Phi(X)
	=
		\Phi_0(X)
		+
		\ell n^{-2\ell} (2\gamma Q_2)^{4\ell} \!\:O\!\!\;\Biggl(\!\Lambda^2\!\!\; \sum_{i=1}^{\ell-1} (2\gamma Q_2)^{-i}
	&+
		\Lambda \!\sum_{i=\ell}^{2\ell-1}	(2\gamma Q_2)^{-i}\!
	\\&+
		\ell^{-1} (2\gamma Q_2)^{-2\ell}\Biggr).
\end{aligned}
\!\!\!
\end{equation}
For $\ell \in \omega(1)$, the asymptotic expression of the previous equation is bounded from above by $O(\Lambda^2)$, provided that $\poly(\ell) (2\gamma Q_2)^{-\Theta(\ell)} \subset o(1)$.
For the latter to hold, it suffices that ${2\gamma Q_2 - 1 \in \omega\bigl(\ell^{-1} \log(\ell)\bigr)}$.
We then obtain the upper bound
\begin{align}
		\Phi(X) \; =\;
		\Phi_0(X)
		\,&+\,
		O\Bigl( \ell \Lambda^2 n^{-2\ell} (2\gamma Q_2)^{4\ell} \Bigr).
\end{align}
We may show that $\Phi(X) = \Phi_0(X) \bigl[1 + o(1)\bigr]$: using Eqn.~\eqref{eqn:permutationApprox}, we may estimate
\begin{equation}
\mspace{-20mu}
\begin{aligned}[b]
		F_{2\ell-1}
	\;&=\;	
	n \cdot \frac{1}{2}\left[ \tfrac{1}{2} \binom{n-1}{\ell-1} (\ell-1)! \right]
			\left[ \tfrac{1}{2} \binom{n-\ell}{\ell-1} (\ell-1)! \right]
	\\[1ex]&=\;
		\frac{n!}{8(n-2\ell+1)!}
	\;\sim\;
		\tfrac{1}{8} n^{2\ell-1} 
	,\;
\end{aligned}
\mspace{-40mu}
\end{equation}
so that we have
\begin{equation}
\begin{aligned}[b]
		\Phi_0(X)
	\;&\lesssim\;
		F_{2\ell-1} \left(2 \gamma \:\!\big/\;\!\! n \right)^{\!4\ell} Q_2^{4\ell-4} Q_{\text{crux}}^2 
	\;=\;
		\left(\frac{Q_{\text{crux}}^2}{8 Q_2^4}\right) n^{-2\ell-1} \left(2 \gamma Q_2 \right)^{\!4\ell},
\end{aligned}
\end{equation}
whereas by $\ell \in o(n^{1/9})$ and $\Lambda \in \Theta(\ell^4/n) \subset o(n^{-5/9})$ we have
\begin{equation}
  O\Bigl(\ell \Lambda^2 n^{-2\ell} (2\gamma Q_2)^{4\ell}\Bigr) \;\subseteq\; o\Bigl(n^{-2\ell-1} (2\gamma Q_2)^{4\ell}\Bigr) .
\end{equation}
We then have $\Phi(X) \sim \Phi_0(X)$ as promised.
Thus we have $\bE(\varphi_\ell^2) \sim \bE(\varphi_\ell)^2$, so that $\Var(\varphi_\ell) \in o(\bE(\varphi_\ell)^2)$.
By Chebyshev's inequality, the probability that $\varphi_\ell^2$ varies from its mean by $\omega(\Var(\varphi_\ell))$ is zero; then in particular $\varphi_\ell$ is almost surely greater than $1$ provided that $\bE(\varphi_\ell) > 1$.

Frustrated subsystems may be efficiently detected when they are present, as follows.
For each vertex $x \in V(G)$, constraint-pair $(\bra{\alpha_h}, \bra{\alpha_j})$, and $\ell > 1$, we may enumerate the number of alternating paths (in the terminology of Ref.~\cite{JWZ11}) of length $\ell$ which begin an end at $x$ whose first constraint is of the form $\bra{\alpha_h} \ox \bra{\gamma}$ and whose final constraint is of the form $\bra{\beta} \ox \bra{\alpha_j}$.
We may do so by traversing all alternating paths starting at $x$ by a breadth-first search, and noting at each step whether in one step we may reach a visited vertex which could be used to close an alternating path back to $x$.
Any one such path represents an alternating or quasi-alternating loop at $x$.
If for any $\ell > 1$ there are two such loops with inconsistent constraints, then the constraints at $x$ are unsatisfiable.
Exploring all of the alternating paths from $x$ for any one constraint pair $(\bra{\alpha_h}, \bra{\alpha_j})$ can be done in time $O(m)$; doing so for all constraint-pairs and all $x \in V(G)$ can be done in time $O(nmf^2)$.
The frustrated pair of constraints may not represent a frustrated figure eight (\eg~if the alternating paths starting and ending at $x$ are of different lengths), but nevertheless serve to certify that the instance of \num\QSAT[2] is frustrated, and are present for all frustrated instances.

Thus for $m \ge \frac{1 + \epsilon}{2Q_2}n$ for positive $\epsilon \in \omega(n^{-1/9} \log(n))$, an instance of \QSAT[2] constructed on $G$ selected according to the \ErdRen\ distribution will be frustrated almost surely, due to the presence of multiple frustrated figure-eight subsystems of size $O(\poly(n))$.
Furthermore, one may determine that such frustrations exist in polynomial time, when they are present.

\subsubsection{The highly decoupled phase in frustration-free \ErdRen\ models}
\label{sec:critFrustFreeErdRen}

\smallskip\noindent
In constructing frustration-free instances of \QSAT[2] from a discrete distribution, we may suppose that constraints are repeatedly sampled for each new constraint until we obtain one which does not render the instance unsatisfiable.
Any constraint which on the first ``try'' would have resulted in a frustrated instance, we call a \emph{would-be} frustration.
We may then consider the structures in the Hamiltonian which \emph{would have} arisen, had we taken the constraint which was first selected for any interaction, and thus speak counterfactually of such features as ``would-be'' frustrated figure-eight subsystems.

In frustrated figure-eight subsystems $X = X_1 \union X_2$, the common qubit $x_\ell$ has conflicting constraints imposed on it by the two cycles $X_1$ and $X_2$.
If we condition on frustration-free instances, this becomes a \emph{would-be} frustrated figure-eight.
As $X$ is being constructed, one of the cycles (without loss of generality, $X_1$) must be completed before the other: this is either a loop or quasi-alternating loop at $x$ (in the terminology of Ref.~\cite{JWZ11}).
A quasi-alternating loop at $x$ fixes the state of $x$, which by construction do not by themselves satisfy the constraints imposed on $x$ by $X_2$. 
Similar remarks apply when $X_1$ is an alternating loop, which allows two possible single-qubit states for $x$ which on their own satisfy the constraints imposed by $X_1$.
In the case that $X_1$ is an alternating loop, $x$ may be in one of two states $|\psi_x^0\rangle$ or $|\psi_x^1\rangle$ in a product with the rest of $X_1$, in which case all of the other spins of $X_1$ are in a product state $|\Phi^0\rangle$ or $|\Phi^1\rangle$ (respectively) determined by that state, or it may be entangled with the rest of the loop in some superposition $u_0 |\psi_x^0\rangle|\Phi^0\rangle + u_1 |\psi_x^1\rangle|\Phi^1\rangle$.
In either case, the marginal of any satisfying state on $x$ is a mixture of $|\psi_x^0\rangle$ or $|\psi_x^1\rangle$, neither of which on their own satisfy the constraints imposed by $X_2$ on $x$.
Then in any case, upon the completion of the cycle $X_1$, the states of all qubits in $X_2$ which are accessible from $x$ at that time are uniquely fixed.
Each subsequent edge of $X_2$ which connects more qubits to $x_\ell$ also fixes the state of those qubits.
This means in particular that every one of the $\ell$ qubits $v \in V(X_2)$ have fixed states $\ket{\bar\psi_v}$.
We call such a subsystem of fixed qubits a \emph{frozen} subsystem.
Thus, a would-be frustrated figure-eight on $2\ell-1$ qubits contains an (actually) frozen cycle of $\ell$ qubits.

The analysis of the preceding section concerning frustrated figure-eight subsystems $X = {X_1 \union X_2}$ can be used to demonstrate the the existence of a ``frozen core'', or a subgraph of the giant component which itself contains $\Omega(n)$ vertices.
The growth of this frozen core will gradually start to obstruct long-range constraints within the giant component, until eventually it renders the \num\QSAT[2] problem highly decoupled.

To describe the growth of large frozen subsystems in frustration-free \ErdRen\ models, we consider a random graph model for qubits with fixed states.
Define a directed graph $F$ defined by the \QSAT[2] instance consisting of frozen subsystems, including only vertices representing qubits with fixed states, and with arcs $x \rightarrow y$ for qubits connected by constraints $\bra{\eta_{x,y}}$ such that $\bra{\eta_{x,y}} \bigl(\ket{\smash{\bar\psi_x}} \ox \idop\bigr) \ne \vec 0\herm$.
We call this digraph the \emph{frozen subgraph} of $G$.

We may establish lower bounds on the growth of $F$ in terms of an \ErdRen\ graph $U$, where edges of $G$ belong to $E(U)$ independently with some probability $\tilde Q \le Q$, and where all edges of $U$ are covered by arcs of $F$. 
We consider $Q_\infty = 1 - \| \vec q \|_\infty$, and let $p_\infty = mQ_\infty/\binom{n}{2}$.
We then let $U$ be an \ErdRen\ graph having $m_\infty \approx \binom{n}{2} p_\infty$ edges: we treat this as a subgraph of the \ErdRen\ interaction graph $G$,\footnote{%
	We may simulate randomly sampling over graphs with $m$ edges, by considering graphs in which edges are present \iid\ with probability $p = m/\binom{n}{2}$ ---
	the $\sqrt n$ variance in the number of edges is smaller than the scales at which phase transitions such as the emergence of the giant component occur.
}
including each edge of $G$ with probability $Q_\infty$.
Consider a random colouring $c: V \to \{1,2,\ldots,f\}$, in which $\Pr[c(x) = j] = q_j$.
For a given qubit $x$ which has a fixed state $\ket{\smash{\bar\alpha_{c(x)}}}$, and a newly added edge $xy \in E(G)$, the probability that $x \arc y$ is an arc of the frozen subgraph $F$ is $1 - q_{c(x)} \ge Q_\infty$.
From an initial set $S$ of fixed qubits, we then simulate the construction of $F$ as follows:
\begin{enumerate}
\item
	For each newly included vertex $x \in V(F)$ or $x \in S$, assign its colour $c(x)$;
\item
	For each neighbour $y$ of $x$ in $G$: 
		If $xy \in U$, include $x \arc y$ in $F$; otherwise include $x \arc y$ in $F$ with probability $(q_1 - q_{c(x)})/q_1$; otherwise exclude it.
\item
	Repeat the above until all $x \in S$ have been traversed, and no new vertices have been included in $F$.
\end{enumerate}
This construction reproduces the probability distribution of arcs in $F$, with the random colouring of the vertex $c(y)$ taking the place of the action of constraints $\bra{\eta_{x,y}} = \bra{\beta}_x \ox \bra{\alpha_{c(y)}}$ which fixes the state of the qubit $y$.

From the above, we may show that the largest (weakly connected) component of $F$ grows at least as quickly as that of the \ErdRen\ graph $U$ having $m_\infty \sim mQ_\infty$ edges.
In particular, if $\frac{m}{n} > \gamma_\infty$ for $\gamma_\infty := \frac{1}{2Q_\infty}$, then $U$ has a giant connected component $\Gamma\sur{U}$; if any vertices of $\Gamma\sur{U}$ are in $F$, then the entire component $\gamma\sur{U}$ is a subgraph of $F$.
As we have noted, there are frozen cycles (arising from would-be frustrated figure eights) of size $\ell \in \poly(n)$ for $(1+\epsilon)/2Q_2 \le \frac{m}{n} \le \gamma_\infty$: and almost surely a constant fraction of these vertices are subsumed into $\Gamma\sur{U}$, which has size $O(n)$.
Then for $\frac{m}{n} > \gamma_\infty$, the giant component of $U$ is almost surely contained in some weakly-connected component of $F$.
Thus $F$ almost surely contains a frozen core $\Gamma\sur{F}$ for $\gamma > \gamma_\infty$, which is at least as large as $\Gamma\sur{U}$.

Because the qubits in the frozen core cannot mediate non-trivial long-range constraints between non-fixed qubits, and do not contribute to the value of the \num\QSAT[2] instance, they in effect play no role in the solution and may be removed.
Let $\gamma = \frac{m}{n}$.
By Ref.~\cite[Theorem~9b]{ErdRen60}, the subgraph $\Gamma\sur{U}$ contains $(1 - \tfrac{1}{2\gamma Q_{\!\!\:\infty}\!}\xi(\gamma Q_\infty)) n + o(n)$ vertices, where
\begin{equation}
	\label{eqn:xiTaylor}
  \xi(\rho) = \sum_{k\ge1} \frac{k^{k-1}}{k!} (2 \rho\e^{-2 \rho})^k
\end{equation}
and where $\tfrac{1}{2\rho}\xi(\rho)$ expresses (almost surely and up to $o(1)$ error) the fraction of vertices which are contained in tree components in an \ErdRen\ graph with $\rho n$ edges.
Following Ref.~\cite[Theorem~4b]{ErdRen60}, the function $\xi: [0,\infty) \to [0,1]$ has the property that 
$
	\label{eqn:xiDefinition}
  \xi(\rho) \e^{-\xi(\rho)} = 2\rho \e^{-2\rho}
$. 
We may 
show that for any super-critical edge-density $\rho > \tfrac{1}{2}$, there is a sub-critical edge-density $\tilde\rho := \tfrac{1}{2}\xi(\rho) < \tfrac{1}{2}$ such that the distribution of the sizes of tree-components for the edge-densities $\rho$ and $\tilde\rho$ are the same up to a normalization factor.\footnote{%
	Consider a randomly selected tree component $T$, and let 
	$\tau_\rho(t) = \tfrac{1}{2\rho t!} t^{t-2} (2\rho \e^{-2\rho})^t$.
	The probability $P_\rho(t)$ that $T$ has size $t$, when selecting tree-components from the \ErdRen\ graph with $\rho n$ edges, is then
	$
		P_\rho(t)
	\,\sim\,
		\tau_\rho(t) \big/ {\scriptstyle\sum_k} \tau_\rho(k)
$ 
by Ref.~\cite[Eqn.~2.22]{ErdRen60}.
From $\tilde\rho := \tfrac{1}{2}\xi(\rho)$ and Ref.~\cite[Eqn.~4.4]{ErdRen60} we may immediately see that $P_\rho(t) = P_{\tilde\rho}(t)$ for all $t$.
As all but an insignificant number of vertices are contained in either the giant component or in trees, the two distributions on graphs are indistinguishable.} 
Thus deleting the giant component from the \ErdRen\ graph with density $\rho$ gives rise to a graph indistinguishable from an \ErdRen\ graph with density $\tilde \rho$, albeit on $\tfrac{1}{2\rho}\xi(\rho) n$ vertices.
More generally, deleting the subgraph $\Gamma\sur{U}$ from the graph $G$ yields
a graph indistinguishable from an \ErdRen\ graph on $\tfrac{1}{2\gamma Q_{\!\!\;\infty}\!}\xi(\gamma Q_\infty)n$ vertices, with edge-density given by
	\begin{equation}
	\label{eqn:liquidDensity}
	\begin{aligned}[b]
		\!\tilde\gamma &:=
			\tfrac{1}{2}\xi(\gamma Q_\infty) + \gamma(1-Q_\infty)\biggl[\frac{\xi(\gamma Q_\infty)}{2 \gamma Q_\infty} \biggr]^2
		\!=
			\tfrac{1}{2}\xi(\gamma Q_\infty) + \tfrac{1-Q_\infty}{4 \gamma Q_\infty^2} \xi(\gamma Q_\infty)^2 ,
	\end{aligned}
	\end{equation}
	where the first term accounts for the density of $U \setminus \Gamma\sur{U}$, and the second term accounts for the contribution of edges $e \in E(G) \setminus E(U)$ which are also not incident to $\Gamma\sur{U}$.

As the frozen core $\Gamma\sur{F} \supseteq \Gamma\sur{U}$ grows, the subgraph of $G$ that remains after removing $\Gamma\sur{F}$ becomes more sparse, and eventually becomes highly disconnected.
That is to say, the instance with the frozen subsystems included is highly decoupled.
Note that $\xi(\rho) = 2\rho$ for $\rho \in [0,\tfrac{1}{2}]$, achieving a maximum of $1$ and then decreasing for $\rho \ge \tfrac{1}{2}$.	
It follows that $\tilde\gamma = \gamma$ for $\gamma Q_\infty \le \tfrac{1}{2}$, achieving a maximum of $1/2Q_\infty$ and then subsequently bounded by
\begin{equation}
\begin{aligned}[b]
		\tilde\gamma
	\;\le\;
		\left[\tfrac{1}{2} + \tfrac{1 - Q_\infty}{2Q_\infty}\right] \xi(\gamma Q_\infty)
	\;&\le\;
		\tfrac{1}{2Q_\infty}\xi(\gamma Q_\infty)
	\\&\le\;
		\gamma \e^{\xi(\gamma Q_\infty)} \e^{-2\gamma Q_\infty}
	\;\le\;
		\gamma \e^{1-2\gamma Q_\infty}.
\end{aligned}
\end{equation}
If $2\gamma Q_\infty-\ln(2\gamma) > 1$, we then have $\tilde \gamma < \tfrac{1}{2}$.
In this case ${G \setminus \Gamma\sur{U}}$ becomes subcritical and thus highly disconnected; the same is then true of ${G \setminus \Gamma\sur{F}}$.

Thus for $\gamma$ sufficiently large, frustration-free instances of \num\QSAT[2] almost surely contain a frozen core pervasive enough to cause the problem to be highly decoupled.
It is easy to show that such a frozen core can be easily detected, as well, using the same techniques as described in the preceding section for frustrated figure-eights.
We may detect the existence of alternating and quasi-alternating loops at each vertex $x$ in the graph, and then consider the constraints on $x$ and its neighbours to discover an initial set of frozen spins.
Following this, using a single breadth-first traversal, we may discover the entire frozen subgraph and its largest component in particular.
Discovering the frozen core is therefore possible in polynomial time using standard techniques.

\subsection{Bond-percolated lattice graphs}
\label{sec:discreteProbabilisticPercolated}

The analysis for random \QSAT[2] is much simpler for bond-percolated square or cubic lattices.
In this graph model, we take vertices labelled either $(a,b) \in \{0,1,\ldots,L-1\}^2$ or $(a,b,c) \in \{0,1,\ldots,L-1\}^3$, and connect each pair of vertices which differ by $1$ in a single co-ordinate, independently with some probability $p$.
We let $d$ denote the dimension of the lattice, let $n = L^d$ be the number of vertices and $m \sim dpn$ be the expected number of edges.

The analysis of phase transitions in the difficulty of \num\QSAT[2] for independent factor constraints is simpler for percolated lattices than for \ErdRen\ graphs, as cycles arise in the percolated lattice much more easily and as the degree of each vertex is necessarily bounded.
Furthermore, we only expect the largest components to grow with $n$ if $p$ is greater than a ``percolation threshold'' $p_c$~\cite{Grimmett-1999},\footnote{%
		For $d_2$, we have $p_c = \tfrac{1}{2}$; for $d = 3$, we have $p_c \approx 0.24881$; \emph{c.f.}~Ref~\cite{Grimmett-1999}.
		N.B.~For $d=3$ it is not yet known whether there exists an infinite component when $p = p_c$; this is known not to occur for $d = 2$ or $d \ge 19$, and the same is conjectured for $d=3$~\cite[Section~9.4]{Grimmett-1999}.
	}
in which case the largest component is unique and scales as $O(n)$.
For \num\QSAT[2] with independent factor constraints, this allows one to show:
\begin{itemize}
\item
	\num\QSAT[2] is almost certainly efficiently solvable for any value of $p$, as there are overlapping phases of frustrated and highly disconnected instances, occurring respectively for $p \in \omega(n^{-1/7})$ and $p \le p_c \in O(1)$;
\item
	For frustration-free instances of \num\QSAT[2], provided that $Q_\infty := 1 - \| \vec q \|_\infty > p_c$, there is a transition directly from highly disconnected instances for $p < p_c$ to highly decoupled instances for $p \ge p_c$, due to the emergence of frozen subgraph whose components decouple the system into small non-interacting components (in a way which is similar to, but more straightforward than, the analogous phenomenon in models on \ErdRen\ graphs.)
\end{itemize}
In this Section we outline these results in enough detail to indicate how the results may be shown more completely.
Furthermore, results which are similar in quality could also be shown for any lattice model, depending in practise only on the size of the smallest cycles and the percolation threshold $p_c$ of the lattice.

\subsubsection{Critical thresholds for unconditional percolated lattice models}
\label{sec:critUnconditionalPercolated}

If each edge in a $d$-dimensional rectangular lattice (for $d \in \{2,3\}$) is present independently with probability $p \in o(1)$, then the first components with cycles to emerge as $p$ increases are the ones with the fewest edges.
That is, if the probability of there being a component in $G$ which is isomorphic to a graph $g$ is $\Omega(1)$, then $G$ will contain infinitely many isomorphic copies of any component $g'$ for which $|E(g')| < |E(g)|$.
The first components with cycles to emerge are therefore individual square facets of the lattice, which are almost surely absent for $p \in o(n^{-1/4})$, and present in infinite abundance for $p \in \omega(n^{-1/4})$.

\begin{figure}
		~\hfill
		\begin{minipage}{0.65\textwidth}
		\caption{%
			(\emph{Top:}) An isolated ``domino'' subgraph of a square lattice.
			Dashed lines indicate missing edges incident to the subgraph.
			A domino subgraph in a 3D lattice may also occur with the two cycles meeting at a right angle.
			(\emph{Bottom:})			
			Illustration of the three independent paths between the central qubits of a domino subgraph.
			If the constraints acting on $b$ do so with different tensor factors $\bra{\alpha}, \bra{\alpha'}, \bra{\alpha''}: \C^2 \to \C$ and similarly for the constraints $\bra{\beta}, \bra{\beta'}, \bra{\beta''}: \C^2 \to \C$ acting on $e$, and the path-constraints are all non-zero, then these form an infeasible system of constraints on the states of $b$ and $e$.
			Similar remarks apply for any pair of qubits connected by more than two independent paths.
		}
		\label{fig:frustrated-domino}
		\end{minipage}
		\hfill
		\begin{minipage}{0.25\textwidth}
		\bigskip
	\begin{tikzpicture}[scale=0.83,
		outer sep=-2pt,
		every node/.style={circle,fill=black}
		]

		\coordinate (d) at (1,0);
		\coordinate (e) at (2,0);
		\coordinate (f) at (3,0);
		\coordinate (a) at (1,1);
		\coordinate (b) at (2,1);
		\coordinate (c) at (3,1);

		\foreach \x in {1,2,3} {
			\foreach \y in {0,1} {
				\foreach \dx/\dy in {0.75/0,-0.75/0,0/0.75,0/-0.75}
					\draw [line width=1.5pt, densely dashed, gray] (\x,\y) -- ($(\x,\y) + (\dx,\dy)$);
	}}

		\draw [line width=2pt, blue!85!white] (b) --	(e) -- (d) -- (a) -- (b) -- (c) -- (f) -- (e);

		\node [label=below left:$a$] at (a) {};
		\node [label=below left:$b$] at (b) {};
		\node [label=below left:$c$] at (c) {};
		\node [label=below left:$d$] at (d) {};
		\node [label=below left:$e$] at (e) {};
		\node [label=below left:$f$] at (f) {};


		\coordinate (b) at (2,-2);
		\coordinate (b') at (1.5,-2);
		\coordinate (b'') at (2.5,-2);
		\coordinate (a) at (1,-2.6);
		\coordinate (c) at (3,-2.6);
		\coordinate (d) at (1,-3.4);
		\coordinate (e) at (2,-4);
		\coordinate (f) at (3,-3.4);
		\coordinate (e') at (1.5,-4);
		\coordinate (e'') at (2.5,-4);

		\draw [line width=3pt, blue!85!white] (b) --	(e);
		\draw [line width=3pt, blue!85!white] (b') --	(a) -- (d) -- (e');
		\draw [line width=3pt, blue!85!white] (b'') --	(c) -- (f) -- (e'');

		\node [label=above:$b$] at (b) {};
		\node [label=above left:$b$] at (b') {};
		\node [label=above right:$b$] at (b'') {};
		\node [label=left:$a$] at (a) {};
		\node [label=right:$c$] at (c) {};
		\node [label=left:$d$] at (d) {};
		\node [label=below:$e$] at (e) {};
		\node [label=below left:$e$] at (e') {};
		\node [label=below right:$e$] at (e'') {};
		\node [label=right:$f$] at (f) {};
	  
	\end{tikzpicture}
		\end{minipage}
		\hfill~
	\end{figure}
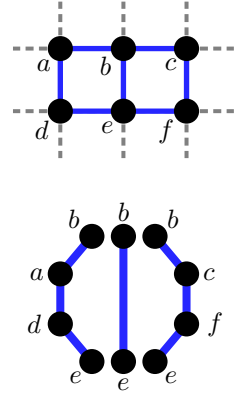

The smallest subgraph of a rectangular lattice which contains two cycles is a \emph{domino graph}, as pictured in Fig.~\ref{fig:frustrated-domino}, which has seven edges.
These are therefore almost certainly absent for $p \in o(n^{-1/7})$, and almost certainly abundantly present for $p \in \omega(n^{-1/7})$.
It is not difficult to show that each of these has a constant probability of being a \emph{frustrated domino}: a system similar to a frustrated-figure eight in which the constraints give rise to unsatisfiable restrictions on the state of the two central qubits.
Consider the three independent paths between the central vertices of a domino subgraph (also depicted in Fig.~\ref{fig:frustrated-domino}).
Given that each edge represents a non-zero constraint (which happens with constant probability), the two outer paths in the domino each give rise to a non-zero path constraint with probability $Q_2^2 = (1 - \| \vec q \|_2^2)^2$.
With some probability, the three path constraints will act on each of their endpoints in a different way from the others.
This remains true even for classical instances of \num\SAT[2], if the constraint-operators are chosen from a probability distribution over a distribution on $\{ \bra{00}, \bra{01}, \bra{10}, \bra{11} \}$ in which each element occurs with probability $\Omega(1)$, each such domino is unsatisfiable with constant probability, in which case the entire instance of \num\SAT[2] which contains it has value zero.
(This would occur, for instance, for an independent factor distribution $\vec q = (q_1, q_2)$ in which $\bra{\alpha_1} = \bra{0}$ and $\bra{\alpha_2} = \bra{1}$, where $q_1$ and $q_2$ are both bounded away from zero.)
Thus, there is a phase transition at $p \in \Theta(n^{-1/7})$ from almost certain satisfiability to almost certain unsatisfiability, due to the probable emergence of frustrated dominoes, of which there are almost surely infinitely many once $p \in \omega(n^{-1/7})$.

The components in a bond-percolated lattice for $p \in O(n^{-1/7})$ almost certainly have size $O(1)$: specifically, they will almost surely have seven vertices or fewer.
Thus the complexity of computing \num\QSAT[2] is almost surely governed by that of multiplying $O(n)$ ``small'' integers.
A simple algorithm to do so is described in Appendix~\ref{apx:multiplyLongList}.
Thus, \num\QSAT[2] is almost surely easy for $p$ increasing up to, and even through, the phase transition at $p \in \Theta(n^{-1/7})$; afterwards, of course, the value is almost surely zero.
Difficult instances of \num\QSAT[2] on percolated lattices are thus either ones which are asymptotically monotone --- that is, for which $Q_2$ decreases with $n$ --- or ones which almost surely never occur.
Similar phenomena will occur for any lattice model, with a phase transition at $p \in \Theta(n^{-1/\beta})$, where $\beta$ is the number of edges in the smallest subgraph having more than one cycle.

\subsubsection{Critical thresholds for frustration-free percolated lattice models}

To obtain interesting instances of \num\SAT[2] or \num\QSAT[2] on a percolated rectangular lattice, 
we must condition on models which are frustration-free.
However, for $p$ less than the percolation probability $p_c$, almost surely the resulting graph $G$ contains only components of size $o(f(n))$ for any $f \in  \omega(1)$.\footnote{%
	For $d = 2$ (for which $p_c = \tfrac{1}{2}$) or $d=3$ (for which $p_c \approx 0.24881$), the distribution of component sizes decreases geometrically (for $p < p_c$)~\cite[Section~6.3]{Grimmett-1999}, and almost surely no infinite component exists.}
This implies that for $p < p_c$, it again suffices to compute the values of \num\QSAT[2] for each component individually,\footnote{%
	As the components all have essentially constant size, this may be done for each component in $O(\log n)$ time, dominated merely by the time required to process the labels of vertices.}
so that \num\QSAT[2] is almost surely efficiently solvable so long as $p \le p_c$.
It thus suffices for us to consider the regime $p > p_c$.

We may proceed similarly to the analysis of the giant component in frustration-free \ErdRen\ models in Section~\ref{sec:critFrustFreeErdRen}.
Would-be frustrated subsystems --- such as frustrated figure-eights on seven vertices (consisting of two square cells intersecting at one qubit) or would-be frustrated dominoes --- will arise in abundance for $p \in \Theta(1)$.
Each one gives rise to several qubits with fixed states, which contribute to the presence of a non-empty frozen subgraph $F$.
If there is a giant component $\Gamma\sur{G}$, then there are almost certainly would-be frustrated subsystems inside it: we ask to what extent these give rise to frozen subsystems which decouple $\Gamma\sur{G}$.

As with the \ErdRen\ case, we may let $Q_\infty = 1 - \| \vec q \|_\infty$ be a lower bound on the probability that any two constraints coinciding at a qubit give rise to a non-zero constraint on a path of length two, such that we may treat this as as independent events even for various pairs of constraints meeting acting on the same qubit.
For instance, the probability that any domino subgraph is a would-be-frustrated domino is at least $Q_\infty^7$.
For any qubit $x \in V(F)$, the probability that some neighbour $y$ in $G$ is also subsumed into $V(F)$ is also at least $Q_\infty$.
We may then consider a percolated lattice model $U$ in which edges are present with probability $Q_\infty$, and any such component which contains a frozen seed gives rise to a component in the frozen subgraph $F$.

When does the frozen core $\Gamma\sur{F}$ decouple an instance of \num\QSAT[2]?
That is: when does ${G \setminus V(\Gamma\sur{F})}$ decompose as a collection of small components?
This relates to the problem, when $U$ has a giant component $\Gamma\sur{U}$, of whether the complement of $\Gamma\sur{U}$ in the complete (square or cubic) lattice has any infinite components (in the limit $n \to \infty$).
For both $d \in \{2,3\}$, there exists a threshold $p_{\text{fin}} < 1$~\cite{GHK-2014} such that the complement of $\Gamma\sur{U}$ in the lattice decomposes into components of finite size when $Q_\infty > p_{\text{fin}}$.\footnote{%
	A simple duality argument shows that $p_{\text{fin}} = p_c = \tfrac{1}{2}$ for $d = 2$~\cite{GHK-2014}.
	For $d = 3$, only know the more general result $p_c \le p_{\text{fin}} < 1$ is currently known.
	While no numerical results are known about $p_{\text{fin}}$ for $d=3$, the growth of infinite clusters in each planar cross-section of the cubic lattice suggests that $p_{\text{fin}}$ is closer to $1-p_c$ than to $1$.
}
Consider the case $Q_\infty > p_c$:
\begin{itemize}
\item 
	If $p = 1$ (that is, $G$ is simply the entire $O(n)$-vertex square or cubic lattice segment), then by construction $G \setminus U$ is a collection of small components.
	As $\Gamma\sur{U}$ is almost surely subsumed by a frozen core $\Gamma\sur{F}$ of qubits with fixed states, which do not contribute to the value of the \num\QSAT[2] instance.
	As the complete lattice with $\Gamma\sur{F}$ removed consists of components of finite size, the resulting instance of \num\QSAT[2] is highly decoupled.

\item
	If $p < 1$, then we may model the resulting \QSAT[2] instance on the percolated lattice by reducing from the previous case (in which the instance is highly decoupled), and removing each constraint in the complete lattice with probability $1-p$: doing so does not make the instance any less decoupled.
\end{itemize}
Thus, for $Q_\infty > p_{\text{fin}}$ (which occurs for $\| \vec q \|_\infty$ below some constant), there is a phase transition for random frustration-free instances of \num\QSAT[2] from highly disconnected instances to highly decoupled instances.
This means that for $d = 2$, difficult instances of \num\QSAT[2] are only likely if the constraint model is ``at least as monotone'' as some distribution of classical \num\SAT[2] constraints; for $d = 3$, a bias towards monotonicity which would be substantial even for \num\SAT[2] is necessary to obtain difficult instances.\footnote{%
	This implies, for instance, that uniformly random \num\SAT[2] on bond-percolated cubic lattices is almost surely efficiently solvable whether or not we condition on satisfiability.}

As a final remark, note that even in the case that $Q_\infty \le p_{\text{fin}}$, there is a chance that frozen subsystems will decouple the largest component $\Gamma\sur{G}$ into small subsystems.
Any domino-shaped subsystem of $\Gamma\sur{G}$ has a finite probability of containing a frozen cycle, which can be treated in the giant component as nodes which are removed from $\Gamma\sur{G}$ with some finite probability $1 - P_{\text{site}} > 0$.
Using results on mixed site- and bond-percolation~\cite{Hammersley-1980}, if ${P_{\text{site}} \,p < p_c}$, the giant component $\Gamma\sur{G}$ still decouples into small subsystems whose degeneracy may be efficiently computed.
We do not present any quantitative results for $Q_\infty \le p_c$, but mention this to indicate that it likely that \num\QSAT[2] may remain easy even for some values $Q_\infty < p_c$, for reasons similar to what we have shown for $Q_\infty > p_c$.

\section{Open questions}
\label{sec:openProblems}

The results of this article may allow for some improvements, which would further bound any ``difficult'' regime in random distributions of \num\QSAT[2] on random graphs.
\begin{itemize}
\item
	For frustration-free instances, $Q_\infty = \min_j (1-q_j)$ is used as a percolation probability on an existing random graph, to obtain lower bounds on the transition to a highly decoupled phase; whereas $Q_2 = \bE_j\bigl[1 - q_j\bigr]$ is used for potentially frustrated models (where we take $\Pr[j] = q_j$).
	Can we replace bounds involving $Q_\infty$ with tighter bounds involving $Q_2$?
\item
	If we remove the condition of frustration-freeness from \num\QSAT[2] altogether, we are left with the problem of computing the degeneracy of the ground-state manifold of a potentially frustrated Hamiltonian.
	Physical intuition suggests that this is typically ``1'', but as with \num\QSAT[2], the classical problem of determining how many boolean strings satisfy a maximum number of constraints is a hard problem in general.
	Under what conditions is it provably easy to compute the ground-state degeneracy of random local Hamiltonians?
\end{itemize}

\subsubsection*{Acknowledgements.}

\smallskip\noindent
This work was partly performed at the University of Cambridge, with support from the EC project QCS.
I would like to thank Ronald de Wolf, as well as an anonymous referee, for helpful comments on the preliminary drafts.


\bibliographystyle{plain}
\bibliography{num-2-qsat}

\appendix


\section{An effective technique for multiplying together long lists of mostly small numbers}
\label{apx:multiplyLongList}

	The value of an instance of \num\QSAT[2] is at most $2^n$.
	We may decompose the value of an instance of \num\QSAT[2] as a product of the values of each connected component.
	In the easily solved instances which arise either when the interaction graph is highly disconnected, or when a large frozen subsystem decouples the Hamiltonian into small independent subsystems, the value of \num\QSAT[2] for these instances is $O(\log n)$.
	One might then show that simply multiplying together these values can be performed in polynomial time, by accounting for the increase in size of the integers involved in the multiplication as more and more factors are included in the product.
	Rather than analyse the growth of the product in an iterative multiplication algorithm, we will show a different algorithm, by which the complexity of evaluating this product is asymptotically no greater than multiplying two $n$-digit numbers.
	
	By sorting the non-giant components of $G$ in order of size (we assume only non-giant components henceforth), we may construct a binary tree such that
	\begin{itemize}
	\item
		The leaves represent sets, each of which contains an individual component and having a stored \num\QSAT[2] value of one more than the component size;
	\item
		Each node which is not a leaf represents the union of the sets of components represented by its child nodes, and stores the product of the \num\QSAT[2] values of its children;
	\item
		The \num\QSAT[2] values of the children of any node are either similar in size (\eg~differing by a factor of at most $3$), or the degeneracy of one of them is constant (\eg~at most $3$).
	\end{itemize}
	We start by pairing the largest component with the second largest component; in the case that the second-largest component is less than half the size of the largest, we first pair it together with a small component (\eg~isolated vertices), and pair the largest component with the parent to these two nodes.
	We continue similarly for the next two largest components, using the smallest components to compensate for differences in the size of the degeneracies of subtrees.
	(Because there are $O(n)$ components in the \ErdRen\ graph for any number of edges $m$, the components of constant size must dominate, and the smallest ones will occur most frequently as a result of the reduced probability of being merged with other components.
	For bond-percolated lattices, the distribution of component sizes is monotone decreasing for any bond-percolation probability $p$, so again small components dominate.)
	The degeneracy of the root node of the tree then is the degeneracy of the Hamiltonian.

	The number of bits required to represent the degeneracy at each level in the tree either remains about constant, or decreases by a factor of $2$, with each level down from the parent node.
	Due to the domination by components of constant size, there will be $\Theta(n)$ leaves on either side of the tree, so that it will have depth $O(\log n)$; most subtrees will be balanced.
	Thus there will be approximately $O(\log n)$ rounds of (in principle parallelisable) multiplications, where the $t\textsuperscript{th}$ round from the final one is between numbers of size $n/2^t$, and each round involves about $2^t$ multiplications in total.
	For any given multiplication algorithm running in some time $O(n^d)$ (\eg~where $d=2$ for the usual straightforward algorithm taught in schools), we can recursively evaluate the value of the entire \num\QSAT[2] instance, corresponding to the root node of the tree, in time
	\begin{equation}
		\begin{aligned}[b]
			\sum_{t=1}^{\mathclap{O(\log n)}}	\, 2^t \left(\frac{n}{2^t}\right)^{\!d}
		&=\;
			\sum_{t=1}^{\mathclap{O(\log n)}}	2^{t(1-d)} n^d
		\,=\;	\left[ \frac{2^{(1-d)} - 2^{O((1-d)\log n)}}{1 - 2^{(1-d)}} \right] \!n^d \;\in\; O(n^d).
		\end{aligned}
		\!\!\!
	\end{equation}

\end{document}